\documentclass[journal]{IEEEtran}
\usepackage{amsmath,amsfonts}
\usepackage{algorithmic}
\usepackage{algorithm}
\usepackage{array}
\usepackage[caption=false,font=normalsize,labelfont=sf,textfont=sf]{subfig}
\usepackage{textcomp}
\usepackage{stfloats}
\usepackage{url}
\usepackage{verbatim}
\usepackage{graphicx}
\usepackage{cite}
\usepackage{booktabs}
\usepackage{tabularx}
\usepackage{booktabs}
\usepackage{tabularx}
\usepackage{braket}
\usepackage{color}
\begin{document}

\title{Full-Wave Green's-Function Modeling of Collective Single-Photon Emission in Non-Markovian Open-System QED with Finite-Bandwidth Compensation of Dispersive Interactions}

\author{Hyunwoo Choi,
        Jisang Seo,
        Junwoo Gim,
        Bowoo Jang,
        Weng C. Chew,
        and~Dong-Yeop~Na
\thanks{H. Choi, J. Seo, J. Gim, B. Jang and D.-Y. Na are with the Department
of Electrical Engineering, Pohang University of Science and Technology, Pohang,
Gyeongsangbuk-do, 37673 Republic of Korea e-mail: (dyna22@postech.ac.kr).}% <-this % stops a space
\thanks{W. C. Chew is with the Elmore Family School of Electrical and Computer Engineering, Purdue University, West Lafayette, IN, 47907 USA.}
}

\markboth{Preprint}%
{Shell \MakeLowercase{\textit{et al.}}: A Sample Article Using IEEEtran.cls for IEEE Journals}

\maketitle

\begin{abstract}
This work presents a full-wave Green's function framework for modeling collective and coherent single-photon emission from multiple quantum emitters embedded in complex electromagnetic structures. Starting from a transverse modal completeness relation of modified Langevin noise formalism, we derive a closed set of coupled equations for population dynamics and frequency-resolved field amplitudes in the single-excitation regime. Since the electromagnetic reservoir is not traced out at the level of the dynamical amplitudes, the emitted single-photon dynamics can be modeled within the same closed set of equations without Markovian approximation in open and dissipative environments.
We demonstrate that finite-bandwidth truncation of the spectral density leads to systematic deviations in coherent dispersive interactions, even when dissipative rates appear converged. To restore causal consistency, we introduce a counter-term compensation scheme that restores the missing dispersive contributions without modifying the retained non-Markovian memory kernel. To validate the scheme and demonstrate the practicality of the proposed framework, we present numerical examples ranging from benchmark configurations to a three-dimensional dispersive ring-resonator structure via finite element method. These capabilities provide a practical route for rigorously incorporating full-wave electromagnetic simulations into non-Markovian multi-emitter quantum electrodynamics, enabling predictive modeling of collective emission, coherent energy exchange, and single-photon radiation in realistic open structures.
\end{abstract}


% \begin{multicols}{2}
\section{Introduction}\label{introduction}

\IEEEPARstart{P}{recise} electromagnetic (EM) modeling is essential for analyzing wave propagation in complex microwave and photonic structures. With the growing integration of quantum emitters and artificial atoms into engineered EM environments for quantum devices (e.g., deterministic single-photon sources~\cite{Nature_quantum_emitter_2021,PI_quantum_emitter_2022,AIP_quantum_emitter_2022,AIP_quantum_emitter_color_center}, coherent quantum interfaces~\cite{2022_quantum_memory,quantum_memory_2023,RMP_cavityQED}, and robust qubit platforms~\cite{Koch_transmon,APM_Transmon_Roth,science_qubit}), their emission dynamics can no longer be characterized independently of the surrounding structure. Instead, the dispersion, radiation leakage, material absorption, and multiple-scattering response directly shape the emitter dynamics and photon-mediated couplings~\cite{gerryandnight,BreuerPetruccione2002_OpenQuantumSystems}. These effects call for a treatment beyond the classic free-space theory of collective spontaneous emission established by Dicke and Lehmberg~\cite{Dicke1954,Lehmberg1970I,Lehmberg1970II}.

Despite the maturity of computational electromagnetics (CEM) techniques~\cite{Jin2014FEM,taflove_fdtd,nature_2023_fdtd}, existing approaches often overlook fine-grained full-wave EM information. Recent unifying viewpoints for non-Markovian simulation, such as quantum dissipation in minimally extended state space (QD-MESS)~\cite{RMP_non_markovian_2026}, show that methods including hierarchical equations of motion (HEOM)~\cite{HEOM_2020_Tanimura,JuliaHEOM_2023,IOHEOM_2025_PRA} and pseudomode models~\cite{2018_pseudomode,2020_pseudomode} can be interpreted as extended-state descriptions of reservoir memory. In these approaches, the nonlocal memory of the reservoir is reproduced by effective or auxiliary degrees of freedom (DoFs), allowing the reduced emitter dynamics to be propagated in an efficient manner. These auxiliary variables, however, are introduced to encode prescribed reservoir spectra rather than to represent the physical DoFs of an actual full-wave structure. Consequently, the emitted photon is not generally retained as a structure-resolved, frequency-dependent dynamical amplitude derived from the full-wave characteristics.
While quasinormal-mode quantization~\cite{QNM_quantization,QNM_review} and few-mode field quantization~\cite{few_mode_quantization_2021,few_mode_quantizatino_2022} have sought to capture the response of complex EM structures within a quantized atom--field framework, they require auxiliary procedures (such as mode selection, quasinormal-mode normalization, or spectral-density fitting) to construct a reduced modal representation.

An alternative route for connecting full-wave EM response to quantum dynamics is macroscopic quantum electrodynamics (QED)~\cite{Scheel2008MacroscopicQED,Gruner1996QEDEvanescent,Philbin_2010}. In this theory, the quantized field is represented in terms of the dyadic Green's function, allowing the dispersive and absorptive response of complex media to be incorporated in a unified manner. However, macroscopic QED introduces Langevin noise sources associated with material absorption. For open radiating structures, this description must be supplemented by fluctuations associated with radiation escaping through the exterior boundary~\cite{stefano,PhysRevA.95.023831}.

More recently, the modified Langevin noise (M-LN) formalism~\cite{Na2023quantumEMLossy,ciatonni_MLN_2024,ciattoni_MLN_2026,choi_2026_dissipative_channel,Hood_MLN} has addressed a limitation of macroscopic QED by placing material absorption and radiative leakage on an equal footing. It decomposes the positive-frequency electric-field operator into medium-assisted (MA) and boundary-assisted (BA) contributions, which describe fluctuations generated inside lossy media and fluctuations associated with the open radiation boundary, respectively.
When the atomic DoFs are coupled to the complete BA--MA polaritonic reservoirs, the M-LN formalism provides a first-principles description of atom--field interactions in open EM systems. Here, first-principles means that the relevant polaritonic DoFs are explicitly retained, so that the effective modes are tracked within a atom-field Hamiltonian framework. In principle, this enables the simultaneous modeling of atomic/photonic observables, including emitted single-photons~\cite{Choi_FDTDQE,Choi_PRApplied,Choi_twoquanta,Forestire_nanophotonics_2025,miano_ECM_PRA_2026}. A direct implementation, however, would require tracking the full continuum of polaritonic reservoir DoFs, making brute-force three-dimensional simulations computationally prohibitive.

In this article, we present a dyadic Green's function formulation for multi-emitter QED based on the modified Langevin noise formalism. The computational tractability of the framework is achieved through two key steps: (i) the dynamics is rigorously modeled in the single-excitation manifold under the rotating-wave approximation, where the total excitation number is conserved; and (ii) the BA--MA transverse modal completeness relation~\cite{Chew_completeness} is used to compress the continuum polaritonic DoFs into the imaginary part of the full-wave dyadic Green's function. This yields a closed set of coupled equations for emitter amplitudes and frequency-resolved single-photon amplitudes, enabling non-Markovian multi-emitter simulations in realistic three-dimensional EM structures without explicitly propagating the full BA--MA mode continuum. Moreover, because the field amplitudes are retained explicitly, the emitted single-photon field is obtained exactly within the same closed set of equations, without invoking the quantum regression theorem~\cite{QRT_1963,BreuerPetruccione2002_OpenQuantumSystems}.

Notably, we identify a finite-bandwidth inconsistency that arises when numerically evaluated Green’s functions are used in non-Markovian quantum dynamics. Even when the retained spectral window reproduces the dissipative memory kernel, the corresponding coherent interactions can remain systematically biased because their causal reconstruction requires spectral information outside the simulated bandwidth. This inconsistency is particularly pronounced for closely spaced emitters and structured photonic environments, where dispersive exchange interactions can be strongly enhanced. Because conventional few-mode descriptions are typically optimized to reproduce a prescribed spectral density, the resulting cutoff-induced bias in the coherent interaction may remain undetected. We resolve this issue by introducing a counter-term compensation scheme that restores the missing dispersive contribution using the independently evaluated real part of the dyadic Green’s function, while leaving the explicitly propagated finite-bandwidth memory kernel unchanged.

We validate the proposed formulation through analytic and numerical simulation examples. A free-space benchmark is first used to verify the counter-term compensation against an analytic coherent exchange rate. We then examine a structured environment with a Drude plasmonic nanosphere to show that finite-bandwidth truncation can leave the dissipative response converged while still producing a sizable dispersive error. Finally, we apply the framework to a three-dimensional dielectric ring resonator using finite-element Green's-function data, demonstrating that the proposed formulation can be coupled to practical full-wave simulations of realistic electromagnetic structures.

The remainder of this paper is organized as follows. In Sec. II, we introduce the dyadic Green's function formulation and derive the coupled equations governing the emitter and frequency-resolved field amplitudes within the single-excitation manifold. In Sec. III, we analyze the analytic structure of the self-energy and present the finite-bandwidth compensation scheme. In Sec. IV, we provide numerical examples illustrating the effect of spectral truncation and the proposed correction. Finally, Sec. V summarizes the main results and discusses possible extensions.

\section{Coupled Equations of Motions for Populations and Single Photon Amplitudes}
Consider a system of $N_a$ atoms located in the vicinity of an open and dissipative electromagnetic (EM) environment.
The environment is assumed to be arbitrary and may consist, for example, of finite-sized absorbing dielectric objects embedded in an open homogeneous host medium such as vacuum.
The $p$-th atom, located at $\mathbf{r}=\mathbf{R}_a^{(p)}$, is characterized by its bare transition frequency $\omega_{a}^{(p)}$ and electric dipole moment $\mathbf{d}_a^{(p)}$.
We focus on situations in which the atoms are initially in their ground states, are excited by external laser fields, and subsequently undergo spontaneous emission induced by vacuum fluctuations.
During this process, the atoms interact not only with the EM environment but also with one another through the self-generated single-photon fields.
Such atom--field and inter-atomic interactions are inherently bidirectional and give rise to complex, non-Markovian dynamics.
To analyze these non-Markovian multi-atom dynamics and to explicitly resolve the spatiotemporal structure of the emitted single-photon fields, we employ the multipolar Hamiltonian combined with the modified Langevin noise formalism.
This formulation is first-principles in the sense that it does not phenomenologically eliminate field DoFs, but instead treats the EM field DoFs explicitly, thereby naturally capturing multi-atom--field interactions in open and dissipative environments.
Within this framework, the field DoFs are represented by Langevin noise operators associated with the environment.
However, a direct coupling of these continuum field DoFs to multiple atoms formally requires an infinite number of modes, rendering brute-force numerical simulations impractical for predictive modeling and device engineering.
To overcome this limitation, we adopt a novel strategy that systematically lumps the continuum degeneracy of the BA--MA field modes at each frequency.
As a result, the number of effective field DoFs is reduced to the number of sampled frequencies, while retaining full information about the emitted single-photons.
The key to this reduction lies in the transverse modal completeness of the BA--MA field modes.
Importantly, we emphasize that omitting the BA-mode contribution, as has been done in several previous studies, formally violates this completeness.
Interestingly, those works nevertheless arrive at correct final expressions because the imaginary part of the dyadic Green's function implicitly encodes the full modal information.
In other words, although the final results are numerically correct, the underlying theoretical procedure is incomplete.
For a consistent and rigorous formulation, the BA-mode contribution must be explicitly included to ensure transverse modal completeness.

\subsection{Modified Langevin noise formalism and transverse modal completeness}

We consider an open electromagnetic (EM) system featuring a lossy medium of finite volume $V_m$ bounded by surface $\partial V$. The macroscopic medium can be characterized by the environment's response via a complex relative permittivity $\epsilon_r(\mathbf{r},\omega)$, which naturally satisfies the Kramers–Kronig relations to ensure causality. In this system, electromagnetic energy dissipates through two distinct channels: (i) ohmic losses within the material and (ii) radiative losses to the far-field. The latter is a direct consequence of the outgoing radiation condition of Maxwell’s equations, where a non-vanishing surface contribution remains at infinity. The M-LN formalism incorporates this radiative term, thereby augmenting the standard macroscopic approach which typically accounts only for ohmic dissipation. Therefore, the EM field is coupled to two distinct, independent reservoirs, requiring the resulting polaritons to be described in terms of two separate BA and MA modes. 

Specifically, the positive-frequency component of the electric field operator is decomposed as
\begin{align}
    \hat{\mathbf{E}}^{(+)}(\mathbf{r},t) = \hat{\mathbf{E}}^{(+)}_{\text{BA}}(\mathbf{r},t) + \hat{\mathbf{E}}^{(+)}_{\text{MA}}(\mathbf{r},t).
    \label{eq:E_BAMA}
\end{align}
where the BA-field contribution is expressed as
\begin{align}
\hat{\mathbf E}^{(+)}_{\mathrm{BA}}(\mathbf r,t)
&=
i
\int_{0}^{\infty} d\omega
\int_{\mathbb S^{2}} d\Omega
\sum_{s\in\{H,V\}}
\nonumber\\
&\quad\times
\mathbf E_{\mathrm B}(\mathbf r;\Omega,s,\omega)
\hat a_{\mathrm B}(\Omega,s,\omega)
e^{-i\omega t}.
\label{eqn:BA-field-def}
\end{align}
and the MA-field contribution is represented as
\begin{flalign}
\hat{\mathbf{E}}^{(+)}_{\text{(MA)}}(\mathbf{r},t)
&=
i
\int_{0}^{\infty} d\omega
\int_{V_m} d\mathbf{r}'
\sum_{\xi \in \{x,y,z\}}
\nonumber \\
&
\mathbf{E}_{\text{M}}(\mathbf{r};\mathbf{r}',\xi,\omega)
\,
\hat{a}_{\text{M}}(\mathbf{r}',\xi,\omega)
e^{-i\omega t}.
\label{eqn:MA-field-def}
\end{flalign}
where $k=\omega/c$ is the free-space wavenumber, $\hat{a}_{\text{B}}$ ($\hat{a}_{\text{M}}$) and $\mathbf{E}_{\text{B}}$ ($\mathbf{E}_{\text{M}}$) denote the annihilation operators and mode functions for the BA (MA) fields, respectively. The geometric parameters include the solid angle $\Omega \in \mathbb{S}^2$, polarization $s$, and dipole orientation $\xi$.
The annihilation and creation operators satisfy the canonical bosonic commutation relations, ensuring the statistical independence of the radiative and absorptive reservoirs:
\begin{align}
\left[
\hat a_{\mathrm B}(\Omega,s,\omega),
\hat a_{\mathrm B}^{\dagger}(\Omega',s',\omega')
\right]
&=
\delta(\Omega-\Omega')\delta_{ss'}\delta(\omega-\omega'),
\\
\left[
\hat a_{\mathrm M}(\mathbf r,\xi,\omega),
\hat a_{\mathrm M}^{\dagger}(\mathbf r',\xi',\omega')
\right]
&=
\delta(\mathbf r-\mathbf r')\delta_{\xi\xi'}\delta(\omega-\omega'),
\\
\left[
\hat a_{\mathrm B}(\Omega,s,\omega),
\hat a_{\mathrm M}^{\dagger}(\mathbf r',\xi',\omega')
\right]
&=0.
\end{align}

The resulting field Hamiltonian is given by the diagonal summation of the energies from both reservoirs as
\begin{align}
\hat H_{\mathrm{field}}
&=
\int_0^\infty d\omega
\int_{\mathbb S^2}d\Omega
\sum_s
\hbar\omega
\hat a_{\mathrm B}^\dagger(\Omega,s,\omega)
\hat a_{\mathrm B}(\Omega,s,\omega)
\nonumber\\
&\quad+
\int_0^\infty d\omega
\int_{V_m}d\mathbf r'
\sum_\xi
\hbar\omega
\hat a_{\mathrm M}^\dagger(\mathbf r',\xi,\omega)
\hat a_{\mathrm M}(\mathbf r',\xi,\omega).
    \label{eq:Ham_field}
\end{align}
In particular, the electric field operator is a linear superposition of excitations arising from fluctuations at the specific locations where dissipation occurs. The BA modes represent fluctuation modes originating from the infinite boundary due to the interaction with the radiative reservoir. Since radiative dissipation is geometrically labeled by the wave propagation direction, the BA modes inherently encapsulate these directional characteristics.
This conceptual framework is strictly analogous to the conventional macroscopic QED description of MA modes, where fluctuations are labeled by the spatial distribution of the lossy medium.
The explicit forms of the BA and MA mode functions are given by
\begin{flalign}
\mathbf E_{\mathrm B}(\mathbf r;\Omega,s,\omega)
&=
\sqrt{\frac{\hbar\mu_0\omega^3}{16\pi^3c}}\,
\lim_{R\to\infty}
\biggl[
4\pi R e^{-ikR}
\nonumber\\
&\quad\times
\overline{\mathbf G}_{E}(\mathbf r,R\hat{\mathbf n};\omega)
\cdot \hat{\mathbf e}_{\Omega,s}
\biggr]\,
\\
\mathbf E_{\mathrm M}(\mathbf r;\mathbf r',\xi,\omega)
&=
k^2
\sqrt{\frac{\hbar}{\pi\epsilon_0}}\,
\sqrt{\mathrm{Im}\,\epsilon_r(\mathbf r',\omega)}
\,
\overline{\mathbf G}_{E}(\mathbf r,\mathbf r';\omega)
\cdot \hat{\mathbf e}_{\xi}.
\label{eqn:E_m_GF}
\end{flalign}
Here, \(k=\omega/c\), \(\hat{\mathbf n}\) denotes the unit vector associated with the solid angle \(\Omega\), and \(\hat{\mathbf e}_{\Omega,s}\) is the corresponding transverse polarization vector. The BA mode is written in terms of the far-field amplitude of the dyadic Green's function, where the factor \(4\pi R e^{-ikR}\) removes the spherical decay and phase accumulation of a source located at \(R\hat{\mathbf n}\) in the limit \(R\to\infty\). The MA mode is generated by a fluctuating source inside the lossy medium and is weighted by \(\mathrm{Im}\,\epsilon_r(\mathbf r',\omega)\).

With these definitions, the BA and MA modes satisfy the transverse modal completeness relation
\begin{align}
&\int_{\mathbb S^2} d\Omega
\sum_s
\mathbf E_{\mathrm B}(\mathbf r;\Omega,s,\omega)
\otimes
\mathbf E_{\mathrm B}^{*}(\mathbf r';\Omega,s,\omega)
\nonumber\\
&\quad+
\int_{V_m} d\mathbf r''
\sum_\xi
\mathbf E_{\mathrm M}(\mathbf r;\mathbf r'',\xi,\omega)
\otimes
\mathbf E_{\mathrm M}^{*}(\mathbf r';\mathbf r'',\xi,\omega)
\nonumber\\
&=
\frac{\hbar\mu_0\omega^2}{\pi}
\mathrm{Im}\,
\overline{\mathbf G}_{E}(\mathbf r,\mathbf r';\omega).
\label{eq:BAMA_completeness}
\end{align}
which has been verified in various literature~\cite{Choi_FDTDQE,ciatonni_MLN_2024,Chew_completeness}.

\subsection{Multi-atom MMTC dynamics in the single-excitation manifold}
\label{subsec:multi_atom_single_quanta_Ew}

In this subsection, we extend the single-emitter formulation to the case of $N_a$ two-level emitters (TLSs) embedded in an open and dissipative EM environment.
Starting from the PZW Hamiltonian, we derive a computationally efficient multimode Tavis--Cummings (MMTC)~\cite{Tavis_Cummings_Model} description in the single-excitation manifold.
As in the single-emitter case, the key step is to eliminate the explicit degeneracy index $\lambda$ associated with the BA--MA mode continuum and express the dynamics solely in terms of the dyadic Green's function $\overline{\mathbf{G}}_E$.

\subsubsection{PZW Hamiltonian for open and dissipative environments}
Within the multipolar (electric-dipole) coupling scheme, the total Hamiltonian is written as

\begin{align}
\hat{H}
&=
\hat{H}_{\mathrm{atoms}}
+
\hat{H}_{\mathrm{field}}
+
\hat{H}_{\mathrm{int}}
\label{eq:H_multipolar_total_final}
\end{align}
The atomic Hamiltonian reads
\begin{align}
\hat{H}_{\mathrm{atoms}}
&=
\sum_{p=1}^{N_a}
\hbar \omega_a
\hat{\sigma}^{(p)}_{+}
\hat{\sigma}^{(p)}_{-},
\label{eq:H_atoms_final}
\end{align}
where $\omega_a$ denotes the bare transition frequency of the $a$-th atom and
$\hat{\sigma}^{(p)}_{\pm}$ are the raising and lowering operators. 
The field Hamiltonian is given by Equation~\eqref{eq:Ham_field} and the interaction Hamiltonian is expressed as 
\begin{align}
\hat{H}_{\mathrm{int}}
&=
-
\sum_{p=1}^{N_a}
\hat{\mathbf{d}}_a^{(p)}
\cdot
\hat{\mathbf{E}}(\mathbf{R}_a^{(p)}),
\label{eq:H_int_dE_final}
\end{align}
Here, the electric field operator $\hat{\mathbf{E}}$ represents not merely free-space photons but is rigorously constructed from BA-MA polaritons as defined in Eq. \eqref{eq:E_BAMA}.
\begin{figure}
    \centering
    \includegraphics[width=1\linewidth]{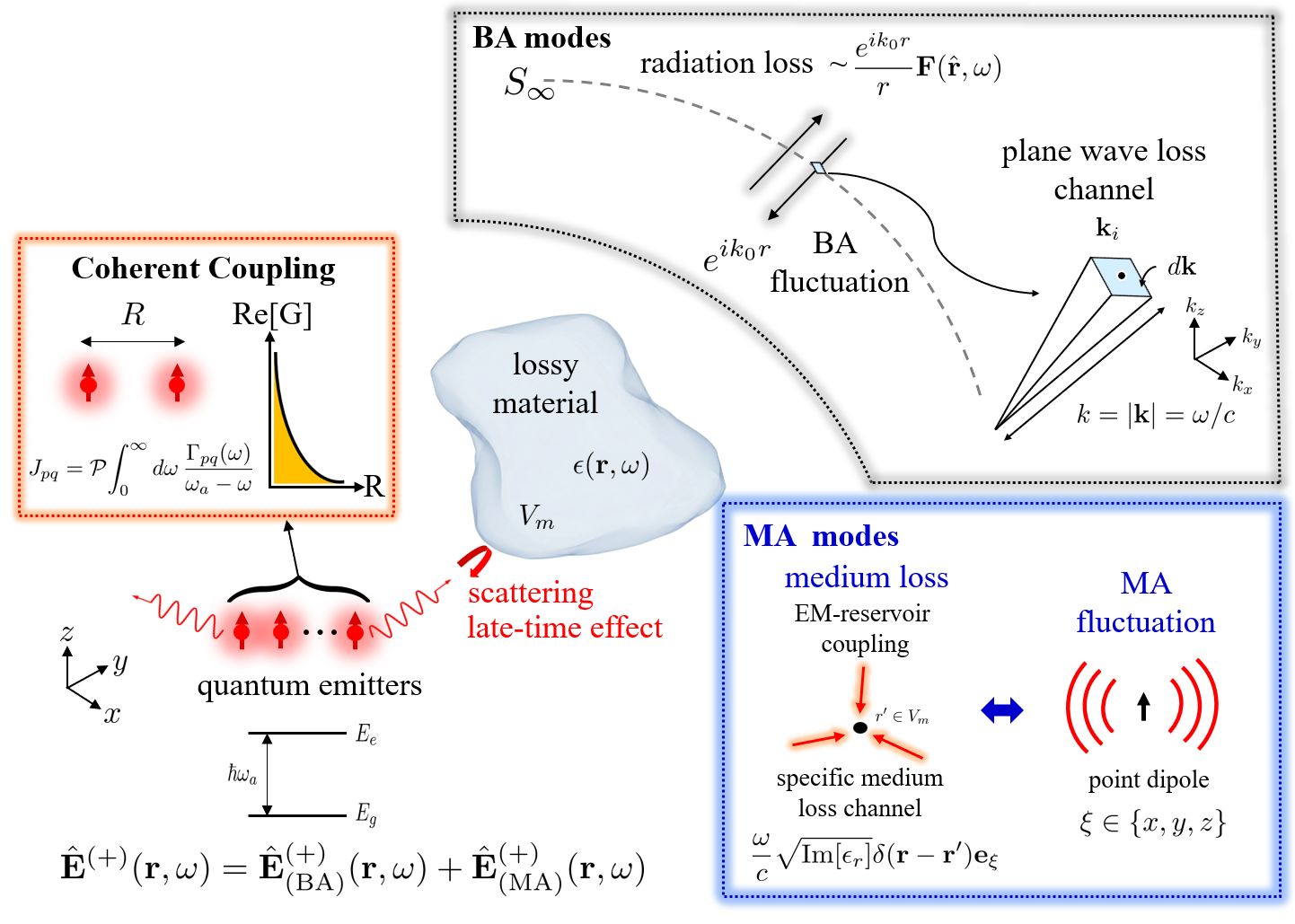}
    \caption{Schematic of the proposed multi-emitter Green's-function formulation. The BA--MA field continuum in an open and lossy electromagnetic environment is compressed into frequency-resolved emitter-coupled amplitudes, enabling single-photon dynamics to be modeled directly from the dyadic Green's function.}
    \label{fig:BAMA}
\end{figure}

\subsubsection{Single-excitation ansatz and amplitude equations}
Using the BA--MA field expansion, we define the atom--field coupling coefficients
\begin{align}
g_{\omega,\lambda}^{(p)}
=
-\,\frac{i}{\hbar}\,
\mathbf{d}_a^{(p)} \cdot \mathbf{E}_{\omega,\lambda}(\mathbf{R}_a^{(p)}).
\label{eq:g_multi_def}
\end{align}
We restrict attention to the single-excitation manifold.
Let $\ket{1_{\omega,\lambda}}$ be the single-photon state in mode $(\omega,\lambda)$ and $\ket{\{0\}}$ the field vacuum.
The most general single-excitation state is
\begin{align}
\ket{\psi(t)}
&=
\sum_{p=1}^{N_a} C_p(t)\,\ket{e_p,\{0\}}
\nonumber \\
&+
\int_{0}^{\infty} d\omega
\int_{\mathcal{D}_{\lambda}} d\lambda.                                                                           
D_{\omega,\lambda}(t)\,
\ket{\{g\},1_{\omega,\lambda}}.
\label{eq:psi_multi_single_excitation}
\end{align}
The state evolves according to the Schr\"odinger equation
\begin{align}
i\hbar\,\frac{d}{dt}\ket{\psi(t)}
=
\hat{H}\ket{\psi(t)}.
\end{align}
Projecting onto $\bra{e_a,\{0\}}$ yields
\begin{align}
i\,\dot{C}_p(t)
&=
\omega_a^{(p)} C_p(t)
+
\int_{0}^{\infty} d\omega
\int_{\mathcal{D}_{\lambda}} d\lambda\;
g_{\omega,\lambda}^{(p)} D_{\omega,\lambda}(t).
\label{eq:Ca_eq_self}
\end{align}
Projecting onto $\bra{\{g\},1_{\omega,\lambda}}$ gives
\begin{align}
i\,\dot{D}_{\omega,\lambda}(t)
=
\omega\,D_{\omega,\lambda}(t)
+
\sum_{p=1}^{N_a}
g_{\omega,\lambda}^{(p)\,*} C_p(t).
\label{eq:Dwlam_eq}
\end{align}

\subsubsection{Elimination of the BA--MA continuum degeneracy}

To eliminate the degeneracy index $\lambda$, we introduce the frequency-projected field amplitudes
\begin{align}
E_{\omega}^{(p)}(t)
=
\int_{\mathcal{D}_{\lambda}} d\lambda\;
g_{\omega,\lambda}^{(p)} D_{\omega,\lambda}(t).
\label{eq:Ew_multi_def}
\end{align}
Following the same procedure as in the single-emitter case, we obtain the closed set of equations
\begin{flalign}
i\,\dot{C}_p(t)
&=
\omega_a^{(p)} C_p(t)
+
\int_{0}^{\infty} d\omega\;
E_{\omega}^{(p)}(t),
\label{eq:MMTC_single_excitation_final_2}
\\
i\,\dot{E}_{\omega}^{(q)}(t)
&=
\omega\,E_{\omega}^{(q)}(t)
+
\sum_{p=1}^{N_a}
\Gamma_{qp}(\omega)\,C_p(t),
\label{eq:MMTC_single_excitation_final}
\end{flalign}
where
\begin{align}
\Gamma_{ba}(\omega)
=
\frac{\omega^2\mu_0}{\pi\hbar}\,
\mathbf{d}_b
\cdot
\mathrm{Im}\!\big[
\overline{\mathbf{G}}_E(\mathbf{R}_b;\mathbf{R}_a,\omega)
\big]
\cdot
\mathbf{d}_a.
\label{eq:Gamma_ba_G}
\end{align}
The diagonal component $\Gamma_{pp}(\omega)$ contains both dissipative and dispersive contributions.

\subsubsection{Single-photon amplitude}

The single-photon electric-field amplitude at an observation point $\mathbf r$ is defined as
\begin{equation}
\mathbf{E}_{\mathrm{spa}}(\mathbf r,t)
=
\langle 0 | \hat{\mathbf E}(\mathbf r) | \psi(t) \rangle .
\label{eq:E_spa}
\end{equation}
Using the BA--MA transverse modal completeness relation, the single-photon field can be expressed solely in terms of the reduced amplitudes $E_{\omega}^{(p)}(t)$ as
\begin{flalign}
\mathbf{E}_{\mathrm{spa}}(\mathbf r,t)
&=
\int_{0}^{\infty} d\omega
\int_{\Omega} d\mathbf{r}'
\mathrm{Im}\!\big[
\overline{\mathbf G}_E(\mathbf r;\mathbf r',\omega)
\big]
\cdot
\mathbf{J}_{\mathrm{spa}}(\mathbf{r}')
\label{eq:spa_multi_final}
\end{flalign}
where
\begin{flalign}
\mathbf{J}_{\mathrm{spa}}(\mathbf{r}')
=
i
\sum_{p=1}^{N_a}\sum_{q=1}^{N_a}
\mathbf d_a^{(p)}
\frac{\mu_0\omega^{2}}{\pi}
\frac{E_{\omega}^{(p)}(t)}{\Gamma_{qa}(\omega)}
\delta(\mathbf{r}'-R_a^{(p)}).
\end{flalign}
Equation~\eqref{eq:spa_multi_final} shows that the complete single-photon radiation pattern in an open and dissipative environment is obtained as a coherent superposition of the contributions generated by each emitter.
All geometric, material, and dissipative properties of the environment enter exclusively through the dyadic Green's function, and no explicit construction of the BA--MA modal degeneracy is required.

\section{Compensation Scheme for Finite Bandwidth Truncation}
\subsection{Causal reconstruction of dispersive interactions}
\label{subsubsec:causal_reconstruction}
While our formulation requires only the dissipative kernel
\(\Gamma_{pq}(\omega)\) as input, the coherent dispersive interactions
(environment-induced Lamb shifts and dipole--dipole couplings) naturally follow
from causal analyticity and the associated dispersion relations
~\cite{Toll1956,Nussenzveig1972}.
We introduce the one-sided Laplace transform in the upper half-plane,
\begin{align}
\tilde{C}_p(z) \equiv \int_{0}^{\infty} dt\, e^{izt} C_p(t),
\end{align}
By applying this to the integro-differential equation for the atomic amplitude derived in Equations.~\eqref{eq:MMTC_single_excitation_final_2} and~\eqref{eq:MMTC_single_excitation_final}, the convolution integral in the time domain transforms into a product in the frequency domain. This yields the following algebraic system for the atomic amplitudes $\tilde{C}_p(z)$,
\begin{align}
(z - \omega_a^{(p)}) \tilde{C}_p(z) - \sum_{q=1}^{N_a} \Sigma_{pq}(z) \tilde{C}_q(z) = i C_p(0),
\label{eq:algebraic_EOM}
\end{align}
where $C_p(0)$ denotes the initial state of the $p$-th atom.
The interactions are now fully governed by the generalized self-energy $\Sigma_{pq}(z)$, which arises from the Laplace transform of the memory kernel
\begin{align}
\Sigma_{pq}(z)
=
\int_{0}^{\infty} d\omega \frac{\Gamma_{pq}(\omega)}{z - \omega}.
\label{eq:generalized_self_energy}
\end{align}

Substituting Eq.~\eqref{eq:Gamma_ba_G} into Eq.~\eqref{eq:generalized_self_energy}
gives the causal self-energy associated with the positive-frequency reservoir
spectrum,
\begin{align}
\Sigma_{pq}(z)
&=
\frac{\mu_0}{\pi\hbar}
\mathbf d_p\cdot
\left[
\int_0^\infty d\omega\,
\frac{\omega^2
\mathrm{Im}\,
\overline{\mathbf G}_E(\mathbf R_p,\mathbf R_q,\omega)}
{z-\omega}
\right]
\cdot\mathbf d_q .
\label{eq:Sigma_Gamma_G}
\end{align}
Taking the boundary value \(z\to\omega+i0^+\) yields
\begin{align}
\Sigma_{pq}(\omega+i0^+)
=
J_{pq}^{\mathrm{PV}}(\omega)
-
i\pi\Gamma_{pq}(\omega),
\end{align}
where
\begin{align}
J_{pq}^{\mathrm{PV}}(\omega_a)
=
\mathcal P
\int_0^\infty d\omega\,
\frac{\Gamma_{pq}(\omega)}{\omega_a-\omega}.
\label{eq:ReSigma_from_Gamma_PV}
\end{align}
The corresponding physical on-shell dispersive interaction is equivalently
obtained from the real part of the dyadic Green's function,
\begin{align}
J_{pq}^{\mathrm{phys}}(\omega_a)
=
\frac{\mu_0\omega_a^2}{\hbar}
\mathbf d_p\cdot
\mathrm{Re}\,
\overline{\mathbf G}_E(\mathbf R_p,\mathbf R_q,\omega_a)
\cdot\mathbf d_q,
\label{eq:J_phys_ReG}
\end{align}
with the diagonal free-space contribution understood to be absorbed into the
renormalized transition frequency. Thus, for \(p=q\), only the scattered
part of \(\mathrm{Re}\,\overline{\mathbf G}_E\) is retained as the
environment-induced Lamb shift.

\subsection{Effects of finite bandwidth truncation}
While the analytic relations theoretically guarantee the reconstruction of dispersive interactions from the full spectral density, the dissipative kernel $\Gamma_{pq}(\omega)$ is inevitably computed over a finite frequency window $\omega\in[\omega_{\min},\omega_{\max}]$. This bandwidth truncation introduces a systematic bias in the reservoir representation. 
First, the time-domain memory kernel is given by
\begin{align}
K_{pq}(t)
\equiv
\int_{0}^{\infty}d\omega\,
\Gamma_{pq}(\omega)e^{-i\omega t},
\qquad t>0 .
\label{eq:memory_kernel_time_domain}
\end{align}
This kernel enters the integro-differential equation for \(C_p(t)\).
Since \(K_{pq}(t)\) is related to \(\Gamma_{pq}(\omega)\) through a
one-sided Fourier transform, the explicitly propagated dissipative dynamics
is controlled by the spectral features of \(\Gamma_{pq}(\omega)\) over the
time scales of interest. Consequently, when the retained spectral window
covers the resonant features that dominate the relevant dynamics, the
finite-band model can reproduce the non-Markovian population dynamics
accurately.

In contrast, the coherent dispersive interaction kernel is obtained from the principal-value (Hilbert-transform) relation, which is intrinsically sensitive to the off-resonant spectral tails of $\Gamma_{pq}(\omega)$. If $\Gamma_{pq}(\omega)$ is known only over the interval $[\omega_{\min},\omega_{\max}]$, the truncated value $J_{pq}^{\mathrm{trun}}$ is expressed as
\begin{align}
J_{pq}^{\mathrm{trun}}(\omega_a)=
\mathcal{P}\!\int_{\omega_{\min}}^{\omega_{\max}} d\omega\,
\frac{\Gamma_{pq}(\omega)}{\omega_a-\omega}.
\label{eq:J_truncation_error}
\end{align}
This truncation introduces a systematic bias,
\begin{align}
\delta J_{pq}(\omega_a)=
\mathcal{P}\!\int_{0}^{\omega_{\min}} d\omega\,
\frac{\Gamma_{pq}(\omega)}{\omega_a-\omega}
+
\mathcal{P}\!\int_{\omega_{\max}}^{\infty} d\omega\,
\frac{\Gamma_{pq}(\omega)}{\omega_a-\omega},
\label{eq:J_truncation_error_2}
\end{align}
which may remain significant even when the corresponding dissipative dynamics is well converged. 

\subsection{Counter-term compensation of finite-band dispersive bias}
\label{subsec:multipolar_hamiltonian_residual}
\begin{figure}
    \centering
    \includegraphics[width=0.9\linewidth]{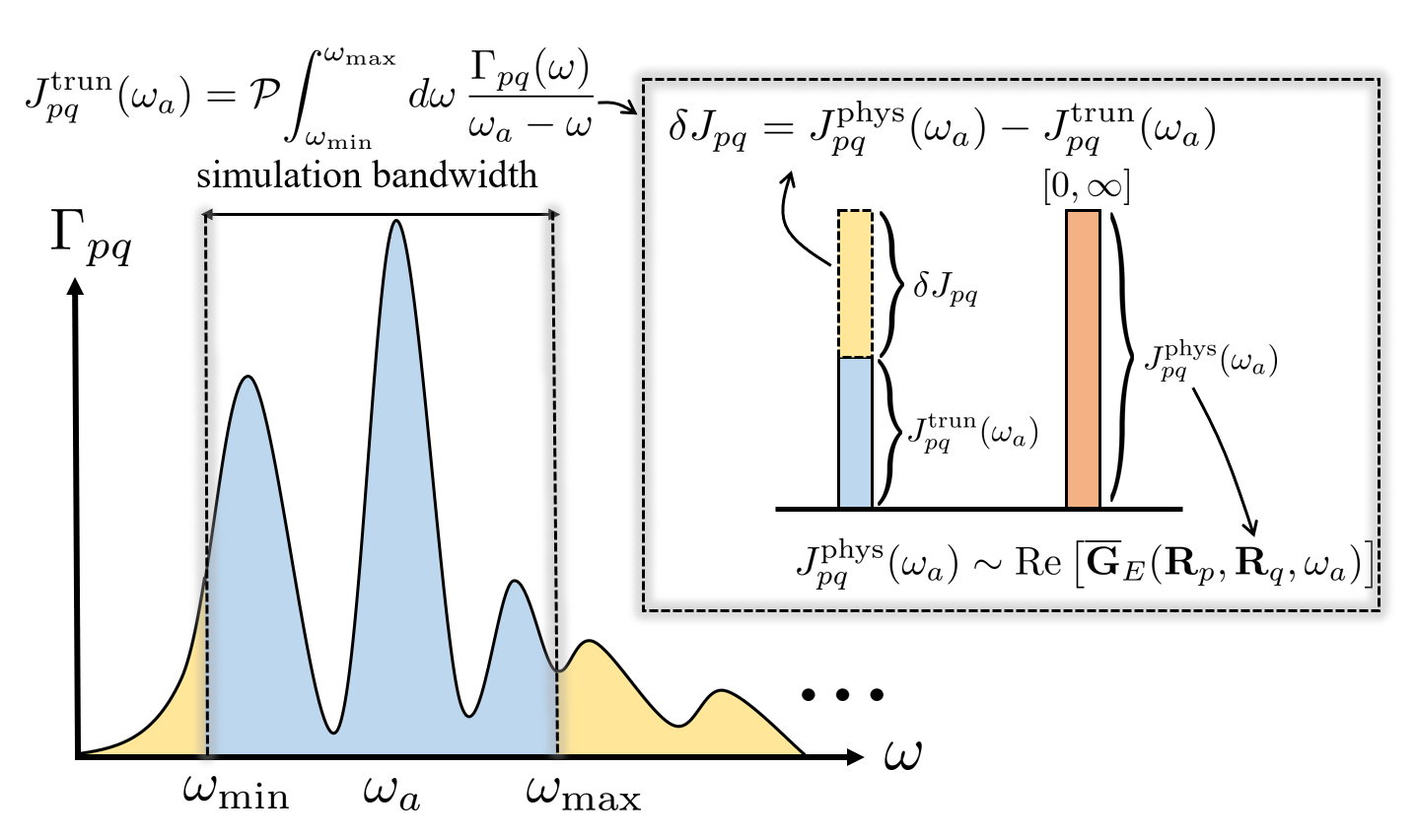}
    \caption{Schematic of the finite-bandwidth compensation scheme. The retained spectral window determines the propagated non-Markovian memory kernel, while the missing out-of-band dispersive contribution is restored through a counter-term Hamiltonian.}
    \label{fig:renormalization_scheme}
\end{figure}

To restore the correct physical dispersive parameters under a finite simulation bandwidth, we introduce a residual (counter-term) Hamiltonian
\begin{align}
\hat{H}_{\delta }
&=
\sum_{p=1}^{N_a}
\hbar \delta J_{pp} \, \hat{\sigma}^{(p)}_{+} \hat{\sigma}^{(p)}_{-}
+
\sum_{p<q}
\hbar \delta J_{pq}\,
\Big(
\hat{\sigma}^{(p)}_{+} \hat{\sigma}^{(q)}_{-}
+
\hat{\sigma}^{(q)}_{+} \hat{\sigma}^{(p)}_{-}
\Big),
\label{eq:H_residual}
\end{align}
This term is added to equation~\eqref{eq:H_multipolar_total_final}, representing the dispersive contributions missing from the finite-bandwidth evaluation of the principal-value integrals.
The residual corrections are defined by subtracting the simulation-generated dispersive shifts from the corresponding physical target values,
\begin{align}
\delta J_{pp}
&=
J_{pp}^{\mathrm{phys}}(\omega_a)
-
J_{pp}^{\mathrm{trun}}(\omega_a),
\label{eq:delta_Lamb}
\\
\delta J_{pq}
&=
J_{pq}^{\mathrm{phys}}(\omega_a)
-
J_{pq}^{\mathrm{trun}}(\omega_a).
\label{eq:delta_J}
\end{align}
Here, the target values ($J_{pp}^{\mathrm{phys}}, J_{pq}^{\mathrm{phys}}$) are determined by the $\textrm{Re}[\overline{\mathbf{G}}]$ from analytic or numerical evaluations. The truncated value $J_{pq}^{\mathrm{trun}}$ is computed via the principal-value integrals (or equivalently, the discrete spectral sums corresponding to the numerical quadrature) over the available frequency window $\omega\in[\omega_{\min},\omega_{\max}]$.
The counter-term prescription corrects the dispersive parameters $(J_{pp}, J_{pq})$ for the bias induced by spectral truncation, while leaving the explicitly retained finite-bandwidth memory effects encoded in $\Gamma_{pq}(\omega)$ unchanged.
Finally, Equation~\eqref{eq:MMTC_single_excitation_final_2} can be rewritten as
\begin{align}
    i\,\dot{C}_p(t)
&=
\bigl[\omega_a^{(p)} + \delta J_{pp}\bigr] C_p(t)
+
\sum_{q\neq p} \delta J_{pq} C_q(t)
\nonumber \\
&+
\int_{\omega_{\min}}^{\omega_{\max}} d\omega\;
E_{\omega}^{(p)}(t),\
\end{align}

\section{Numerical Results}
In this section, we present numerical examples that validate the proposed multi-emitter formulation and demonstrate how the environment-resolved Green's function translates into multi-emitter dynamics. 

\subsection{Verification of the Finite-Bandwidth Compensation Scheme}

We first consider a free-space benchmark, where the analytic dyadic Green's function and the corresponding coherent dipole--dipole interaction are known in closed form. This example provides a direct test of whether the finite-bandwidth simulation recovers the correct principal-value dispersive interaction.
\begin{equation}
    \delta J_{12} = J_{12}^{\mathrm{target}} - J_{12}^{\mathrm{trun}},
\end{equation}
where $J_{12}^{\mathrm{target}}$ is obtained from the analytic free-space Green's function.

We compare three finite-bandwidth simulations. In the uncorrected case, only the explicitly retained spectral window is propagated, so the missing out-of-band dispersive contribution is absent. In the naive-addition case, the full target interaction $J_{12}^{\mathrm{target}}$ is added directly to the finite-bandwidth dynamics. This overcompensates the coherent interaction because the retained spectral window already contains the in-band part $J_{12}^{\mathrm{trun}}$. In the proposed counter-term scheme, only the residual contribution $\delta J_{12}$ is added, so that the retained non-Markovian kernel is left unchanged while the missing dispersive interaction is restored.
The effective coherent exchange rate \(J_{\mathrm{eff}}\) is extracted from the
relative phase evolution of the symmetric and antisymmetric rotating-frame
amplitudes,
\begin{equation}
    C_{\pm}(t)=\frac{C_1(t)\pm C_2(t)}{\sqrt{2}}.
\end{equation}
For the two-emitter exchange Hamiltonian, the symmetric and antisymmetric
branches acquire opposite coherent phases. We therefore extract
\(J_{\mathrm{eff}}\) from the fitted phase difference,
\begin{equation}
J_{\mathrm{eff}}
=
-\frac{1}{2}
\frac{d}{dt}
\arg\left[
\frac{C_+(t)}{C_-(t)}
\right],
\end{equation}
with the sign chosen according to the Hamiltonian convention used in the
time-domain simulation.

\begin{figure}
\centering
\includegraphics[width=0.9\linewidth]{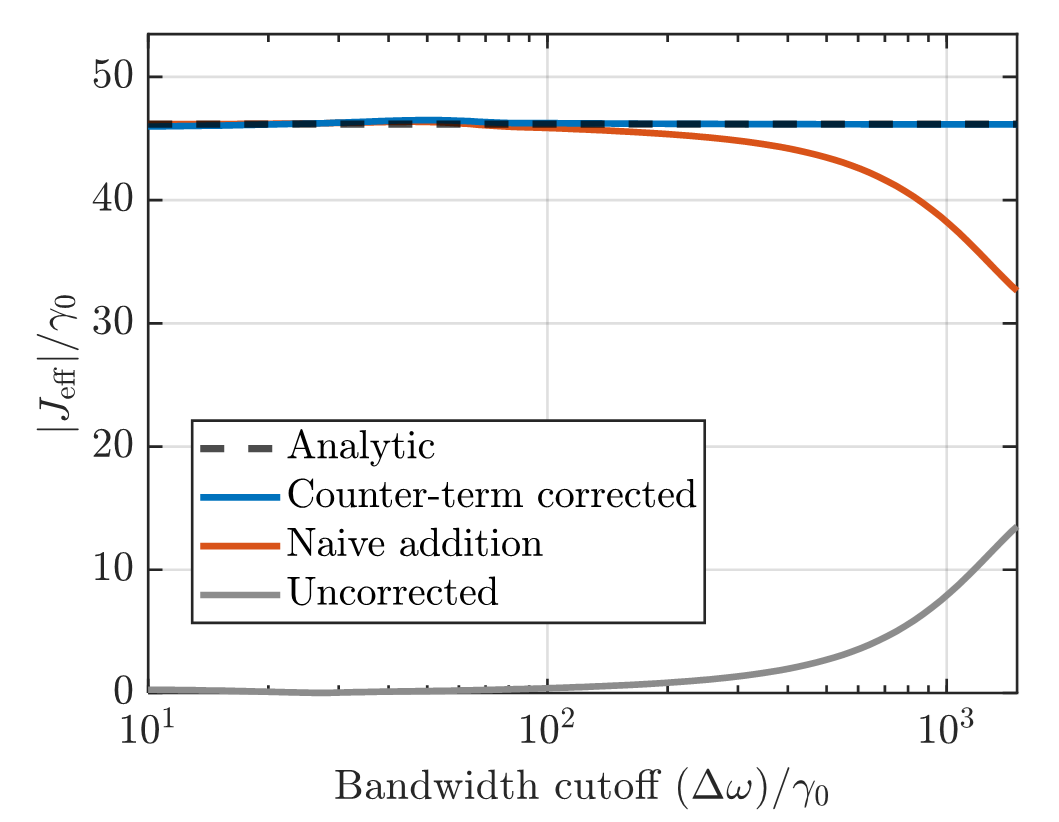}
\caption{Free-space verification of the finite-bandwidth compensation scheme. The counter-term corrected simulation recovers the analytic coherent exchange rate \(J_{\mathrm{eff}}\), whereas the uncorrected and naive-addition cases fail due to missing or double-counted dispersive contributions.}
\label{fig}
\end{figure}

Thus, the comparison tests the dynamical consequence of the compensation scheme. Figure~\ref{fig} shows that the uncorrected simulation fails to recover the analytic interaction over the bandwidth range because the principal-value reconstruction omits high-frequency dispersive components. The naive addition overestimates the interaction by double-counting the in-band dispersive contribution. In contrast, the counter-term corrected result agrees with the analytic ground truth across the simulated bandwidths. This verifies that the proposed correction restores the physical coherent exchange interaction while preserving the explicitly propagated finite-bandwidth memory kernel.

% Figure environment for verification results
We next use the plasmonic nanosphere geometry to separate the behavior of the coherent dispersive interaction from that of the on-shell dissipative response. The nanosphere is modeled by a Drude permittivity
\begin{align}
\epsilon(\omega)
=
1-\frac{\omega_p^2}{\omega^2+i\gamma_p\omega},
\end{align}
where \(\omega_p\) and \(\gamma_p\) denote the plasma frequency and damping rate, respectively. In the numerical example, we set \(\lambda_0=600~\mathrm{nm}\), \(\omega_0=2\pi c/\lambda_0\), \(a=20~\mathrm{nm}\), \(\omega_p=\sqrt{3}\omega_0\), and \(\gamma_p=0.05\omega_0\). The emitters have \(1~\mathrm{D}\) dipole moments oriented along the \(z\)-axis. E1 is fixed \(15~\mathrm{nm}\) away from the sphere surface, corresponding to \(35~\mathrm{nm}\) from the origin, while E2 is displaced along the radial \(y\)-axis by an inter-emitter distance \(R\), swept from \(10~\mathrm{nm}\) to \(500~\mathrm{nm}\). The sphere response is evaluated using the electric-dipole scattering approximation, and the total Green's function is written as \(\mathbf G_{\mathrm{tot}}=\mathbf G_0+\mathbf G_{\mathrm{sc}}\).

The coherent coupling is determined by the real part of the dyadic Green's function, or equivalently by the principal-value reconstruction from the spectral density. Under a finite frequency window, this reconstruction produces a residual
\begin{align}
\delta J_{12}
=
J_{12}^{\mathrm{exact}}
-
J_{12}^{\mathrm{trun}},
\end{align}
where \(J_{12}^{\mathrm{exact}}\) is evaluated from \(\mathrm{Re}\,\mathbf G_{\mathrm{tot}}(\omega_0)\), and \(J_{12}^{\mathrm{trun}}\) is obtained from the truncated principal-value integral over the frequency window \(\omega\in[\omega_{\min},\omega_{\max}]\), corresponding to wavelengths \(\lambda\in[300,900]~\mathrm{nm}\). The integral is evaluated using trapezoidal quadrature on separate frequency grids below and above \(\omega_0\), with the singular point treated by subtracting the on-shell value of \(\Gamma_{12}(\omega_0)\).

\begin{figure}
\centering
\includegraphics[width=0.9\linewidth]{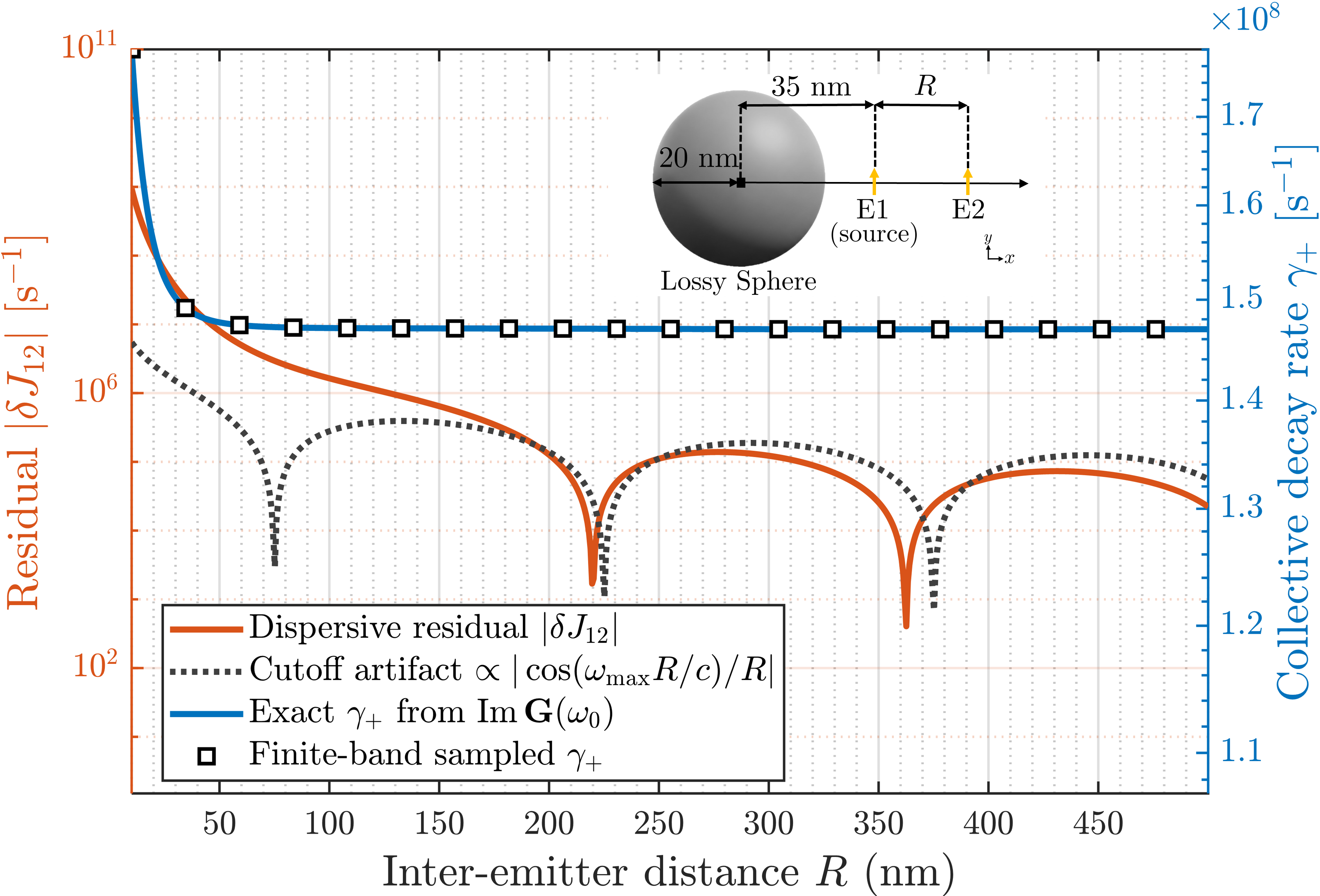}
\caption{Finite-bandwidth truncation in a structured electromagnetic environment. The residual coherent interaction \( |\delta J_{12}| \) exhibits an oscillatory cutoff-induced artifact in the far field and grows strongly at short distances. In contrast, the finite-band samples of the collective decay rate \( \gamma_+ \) overlap with the exact on-shell reference from \( \mathrm{Im}\,\mathbf G(\omega_0) \), showing that the dissipative response is captured while the dispersive interaction remains sensitive to out-of-band spectral contributions.}
\label{fig:nanosphere_bandwidth_response}
\end{figure}
By contrast, the dissipative response at the atomic frequency is determined
locally by \(\mathrm{Im}\,\mathbf G_{\mathrm{tot}}(\omega_0)\). We quantify this
response using the on-shell collective decay-rate matrix
\begin{align}
\gamma_{ij}(\omega_0)
&=
2\pi\Gamma_{ij}(\omega_0)
\nonumber\\
&=
\frac{2\omega_0^2}{\hbar\epsilon_0 c^2}
\mathbf d_i\cdot
\mathrm{Im}\,\mathbf G_{\mathrm{tot}}(\mathbf R_i,\mathbf R_j,\omega_0)
\cdot
\mathbf d_j .
\label{eq:gamma_onshell_nanosphere}
\end{align}
We denote its larger eigenvalue by \(\gamma_+\). The exact reference value of
\(\gamma_+\) is obtained directly from
\(\mathrm{Im}\,\mathbf G_{\mathrm{tot}}(\omega_0)\), while the finite-band sampled
value is obtained by interpolating the sampled
\(\mathrm{Im}\,\mathbf G_{\mathrm{tot}}(\omega)\) values immediately adjacent to
\(\omega_0\).

Figure~\ref{fig:nanosphere_bandwidth_response} shows that finite-bandwidth truncation affects these two quantities in qualitatively different ways. The dispersive residual \(|\delta J_{12}|\) increases rapidly at short inter-emitter distances, where the coherent interaction is dominated by the near-field contribution. In the far-field region, the residual exhibits the expected oscillatory
cutoff artifact, with an approximately \(1/R\) envelope and a phase determined
by the finite spectral boundaries. In contrast, the finite-band samples of \(\gamma_+\) overlap with the exact on-shell reference, confirming that the resonant dissipative response is accurately captured within the finite spectral window. This separation highlights the role of the counter-term Hamiltonian: it restores the missing out-of-band coherent contribution without modifying the explicitly retained finite-bandwidth memory kernel.

\subsection{3D Full-Wave Simulation: Multi-Emitter QED in a Dielectric Ring Resonator}
\label{subsec:ring_resonator_full_wave}

In this subsection, we apply the proposed formulation to a practical three-dimensional dielectric ring resonator containing multiple emitters. The electromagnetic response of the structure is computed using finite-element frequency-domain simulations performed in COMSOL Multiphysics, from which the relevant dyadic Green's-function components are extracted at the emitter positions. The imaginary parts of the Green's function define the frequency-resolved dissipative kernel $\Gamma_{pq}(\omega)$, whereas the real off-diagonal components provide the physical target values for the coherent exchange interaction. In this way, the open and lossy electromagnetic environment enters the quantum dynamics only through the full-wave Green's-function data.

\begin{figure}
\centering

\begin{minipage}{0.85\linewidth}
    \centering
    \includegraphics[width=\linewidth]{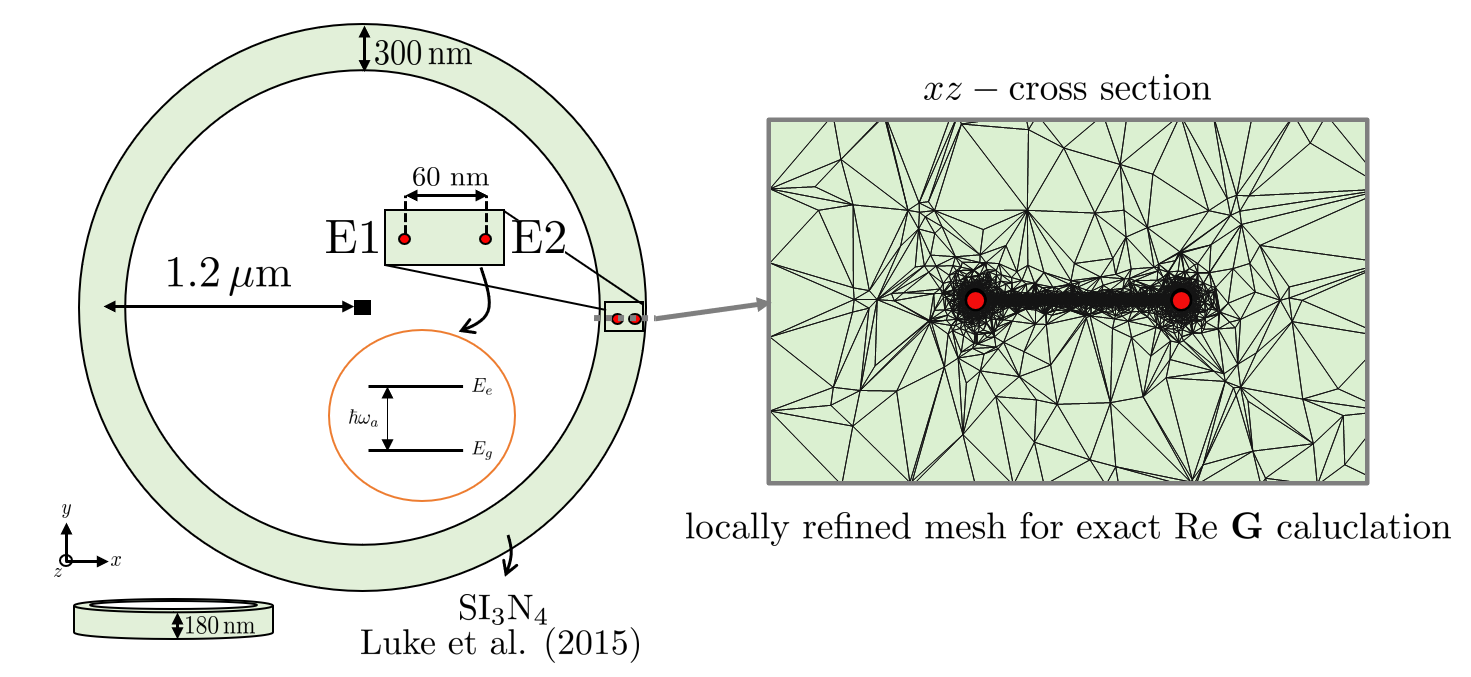}

    {\small (a)}
\end{minipage}
\begin{minipage}{0.48\linewidth}
    \centering
    \includegraphics[width=\linewidth]{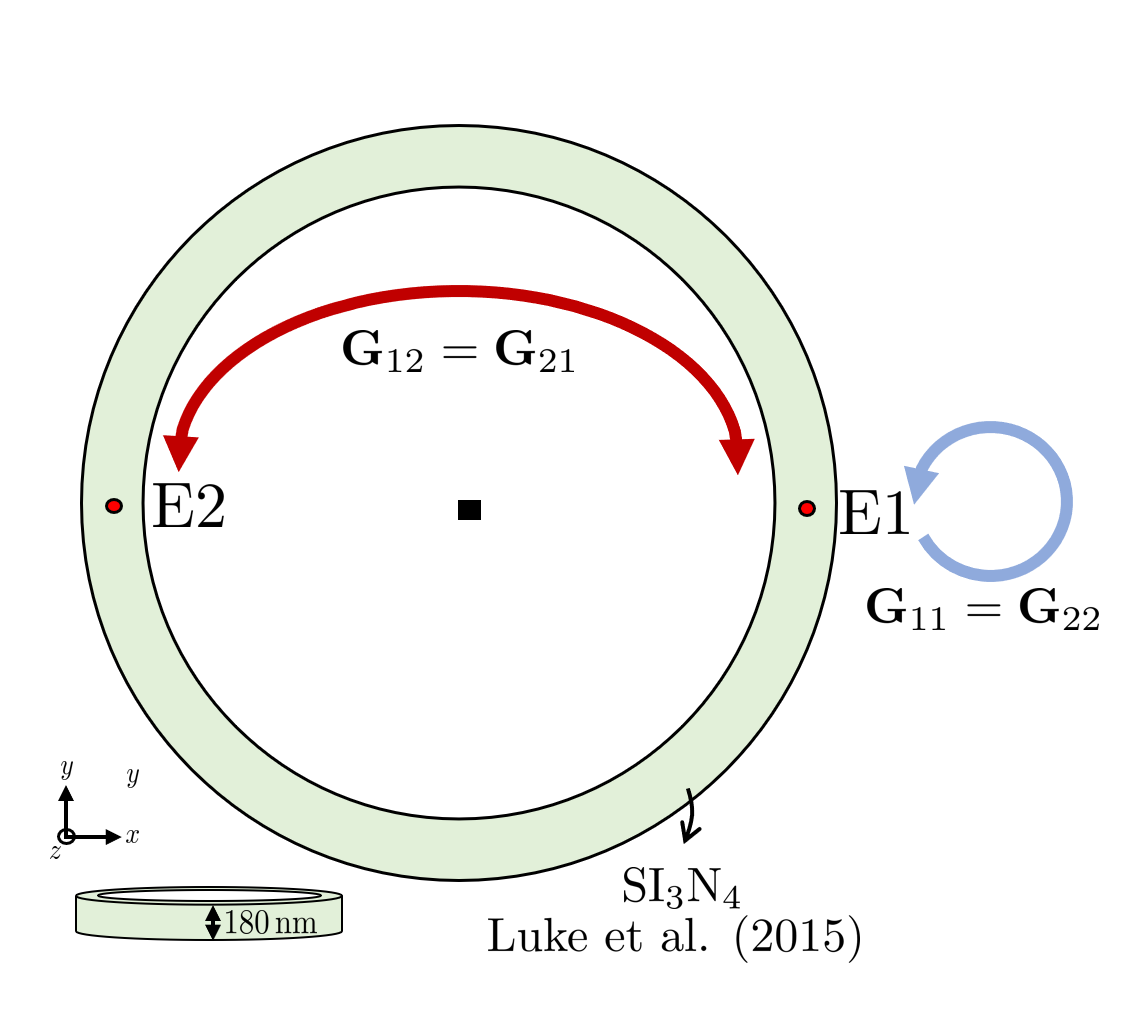}

    {\small (b)}
\end{minipage}
\begin{minipage}{0.48\linewidth}
    \centering
    \includegraphics[width=\linewidth]{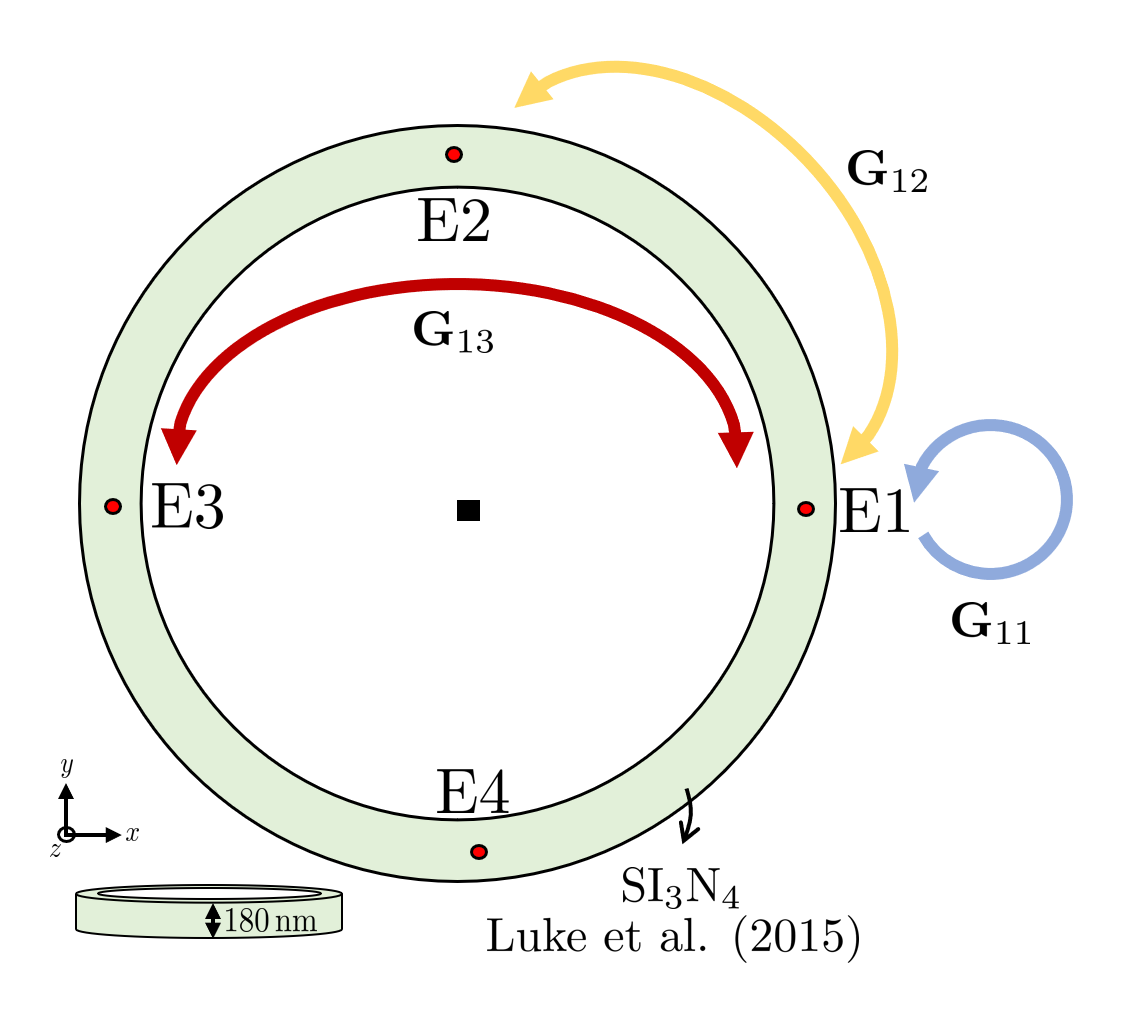}

    {\small (c)}
\end{minipage}

\caption{
Ring-resonator emitter configurations used in the numerical examples.
(a) Close two-emitter configuration.
(b) Distant two-emitter configuration.
(c) Symmetric four-emitter configuration.
}
\label{fig:ring_geometries}
\end{figure}

\begin{figure}[t]
\centering

\begin{minipage}[t]{0.95\linewidth}
    \centering
    \includegraphics[width=\linewidth]{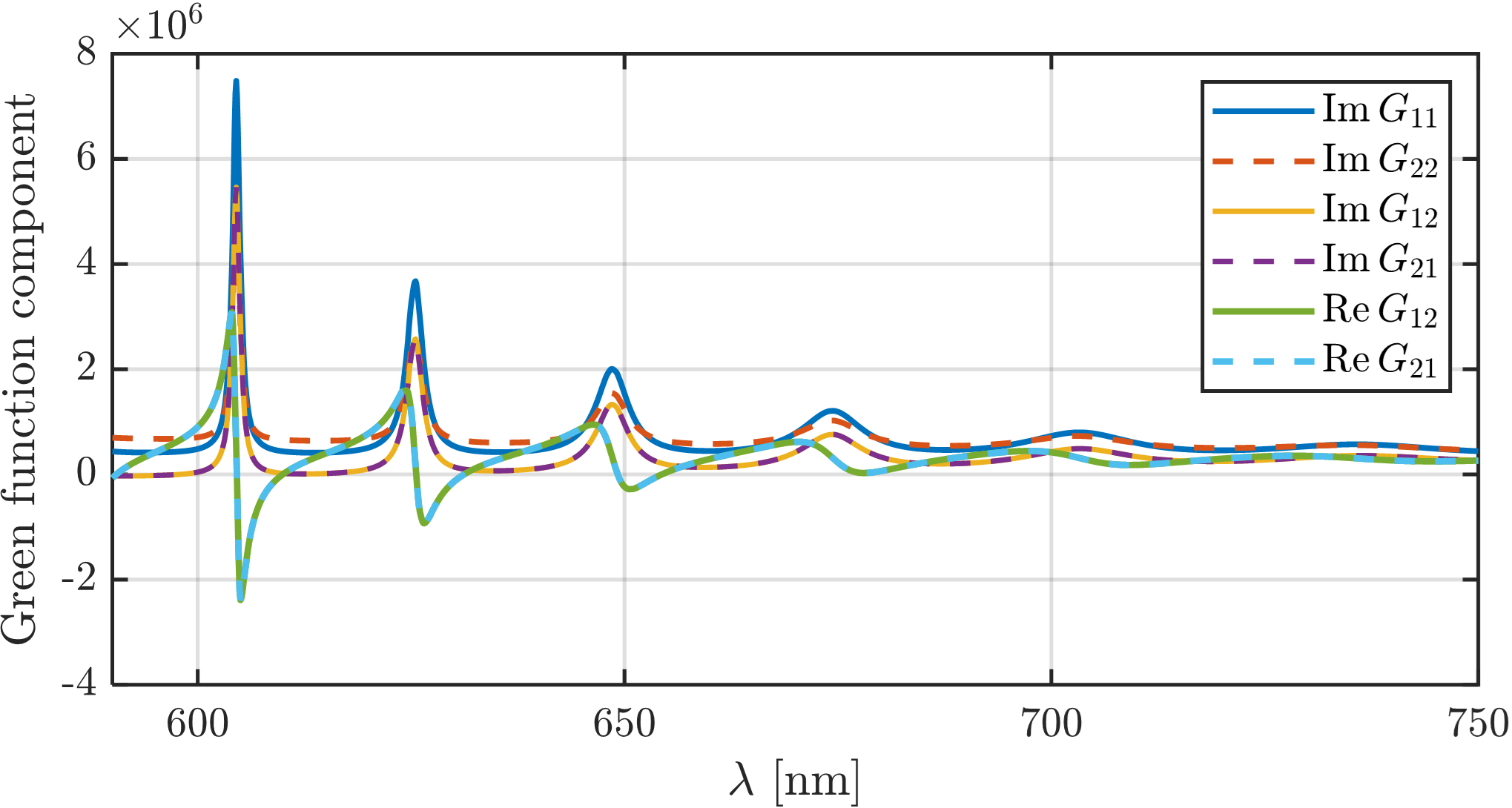}

    {\small(a)}
\end{minipage}

\begin{minipage}{0.95\linewidth}
    \centering
    \includegraphics[width=\linewidth]{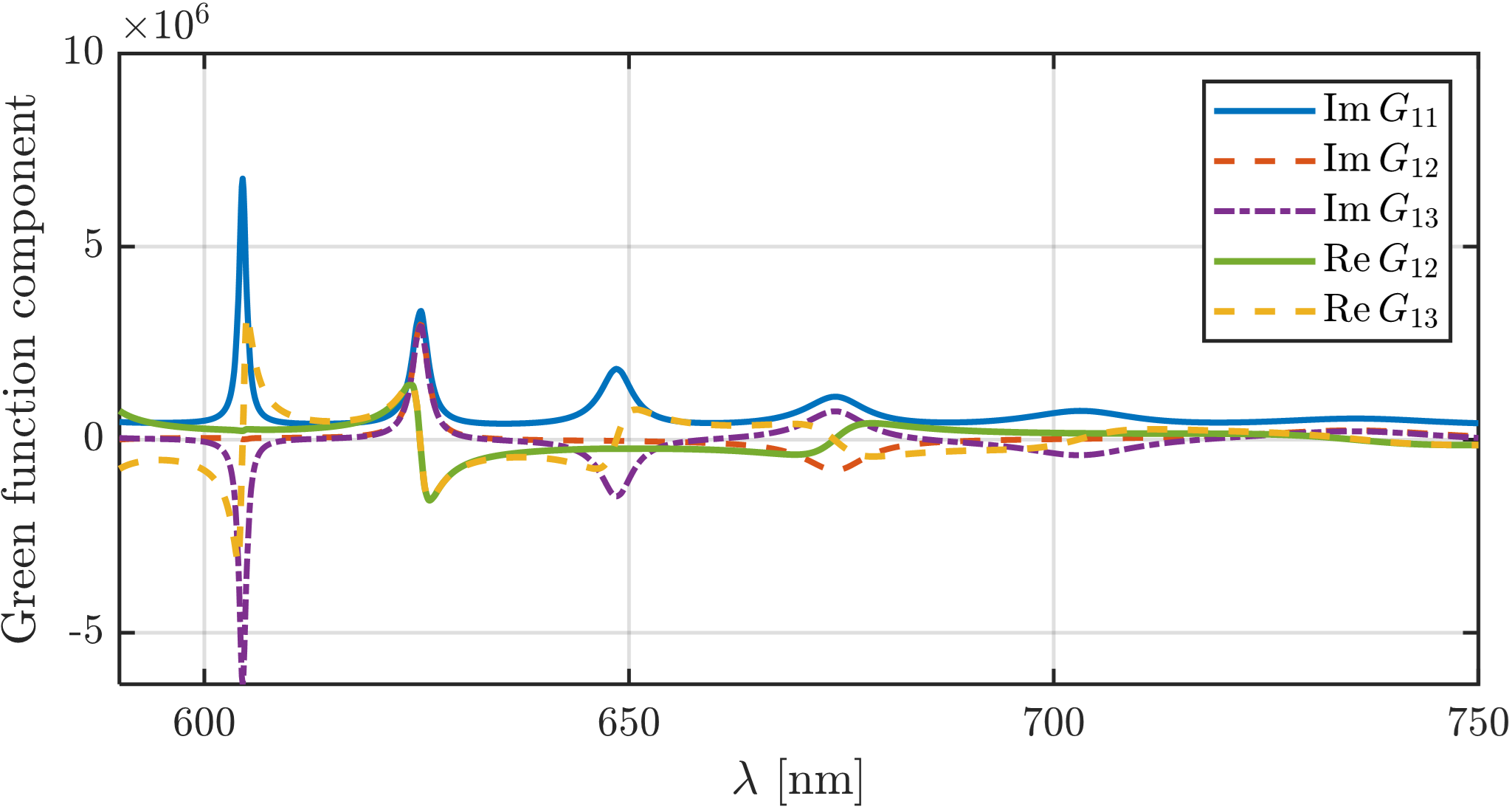}

    {\small(b)}
\end{minipage}
\caption{
Green-function spectra for the ring-resonator emitter configurations.
(a) Close two-emitter case.
(b) Distant two-emitter and symmetric four-emitter case. By symmetry, only the independent components $G_{11}$, $G_{12}$, and $G_{13}$ need to be shown; all other Green-function elements are obtained from these relations. Hence this symmetry-reduced spectrum is used for both Example 2 and Example 3.
}
\label{fig:ring_green_spectra}
\end{figure}

Figure~\ref{fig:ring_geometries} shows the emitter configurations considered in the ring-resonator example. The dielectric ring resonator is modeled using the Si$_3$N$_4$ material dispersion given by Luke \textit{et al.}~\cite{Luke_etal_2015}, and has an outer radius of $1.2~\mu\mathrm{m}$, a waveguide width of $300~\mathrm{nm}$, and a height of $180~\mathrm{nm}$. The electromagnetic Green's functions are obtained from finite-element frequency-domain simulations. 
Figure~\ref{fig:ring_geometries}(a) shows the close two-emitter configuration, where two emitters are placed near the evanescent field region of the ring and separated by $60~\mathrm{nm}$. To accurately resolve the near-field behavior of the dyadic Green's function, especially its real part at the emitter positions, the computational mesh is generated using strong local refinement around the emitters. Figure~\ref{fig:ring_geometries}(b) shows the distant two-emitter configuration, in which the two emitters are placed on opposite sides of the ring. Figure~\ref{fig:ring_geometries}(c) shows the symmetric four-emitter configuration, where four emitters are placed at equally spaced positions around the ring. 
The corresponding Green's-function spectra are shown in Fig.~\ref{fig:ring_green_spectra}. To reduce the number of full-wave Green's-function evaluations, we exploit the geometrical symmetry of the emitter configurations. In the distant two-emitter configuration of Fig.~\ref{fig:ring_geometries}(b), the two emitters are placed at symmetry-equivalent positions on opposite sides of the ring. Therefore, the diagonal responses are identical, $G_{11}=G_{22}$, and the off-diagonal coupling is determined by a single reciprocal component, $G_{12}=G_{21}$. In the symmetric four-emitter configuration of Fig.~\ref{fig:ring_geometries}(c), the rotational symmetry further reduces the independent Green's-function components. All diagonal terms are represented by $G_{11}$, all nearest-neighbor couplings by $G_{12}$, and all opposite-side couplings by $G_{13}$. The remaining matrix elements are obtained from rotational symmetry and reciprocity, rather than being computed independently.

Before turning to the population and single-photon-amplitude dynamics, we
first examine the on-shell collective eigenmode spectrum extracted from the
same full-wave Green's-function data. The effective non-Hermitian interaction
matrix in the rotating frame is
\begin{equation}
\mathbf{H}_{\mathrm{eff}}(\omega_0)
=
\mathbf{J}^{\mathrm{phys}}(\omega_0)
-
i\pi\boldsymbol{\Gamma}(\omega_0),
\label{eq:ring_two_emitter_heff}
\end{equation}
where $\boldsymbol{\Gamma}(\omega_0)$ and $\mathbf{J}^{\mathrm{phys}}(\omega_0)$
follow from the imaginary and real parts of the dyadic Green's function via
Eqs.~\eqref{eq:Gamma_ba_G} and~\eqref{eq:J_phys_ReG}, respectively. The diagonal Lamb shifts
are absorbed into the renormalized transition frequencies, so the branch
splitting is mainly governed by the off-diagonal, resonator-mediated exchange.

The collective eigenfrequencies are obtained from
\begin{equation}
\mathbf{H}_{\mathrm{eff}}(\omega_0)\mathbf{v}_n
=
\epsilon_n(\omega_0)\mathbf{v}_n,
\label{eq:ring_two_emitter_eigenproblem}
\end{equation}
with
\begin{equation}
\epsilon_n(\omega_0)
=
\Omega_n(\omega_0)
-
i\frac{\gamma_n(\omega_0)}{2}.
\label{eq:ring_two_emitter_complex_eigenvalue}
\end{equation}
Here $\Omega_n=\mathrm{Re}\,\epsilon_n$ is the collective dispersive shift in the
rotating frame and $\gamma_n=-2\,\mathrm{Im}\,\epsilon_n$ is the collective
linewidth. The two eigenvectors $\mathbf{v}_n$ describe hybridized collective
states of the emitter pair. For nearly symmetric two-emitter configurations,
they are close to the symmetric and antisymmetric superpositions, while the
corresponding eigenvalues acquire strongly frequency-dependent dispersive
shifts and linewidths near the resonator modes.

\begin{figure}
\centering

\begin{minipage}{0.85\linewidth}
    \centering
    \includegraphics[width=\linewidth]{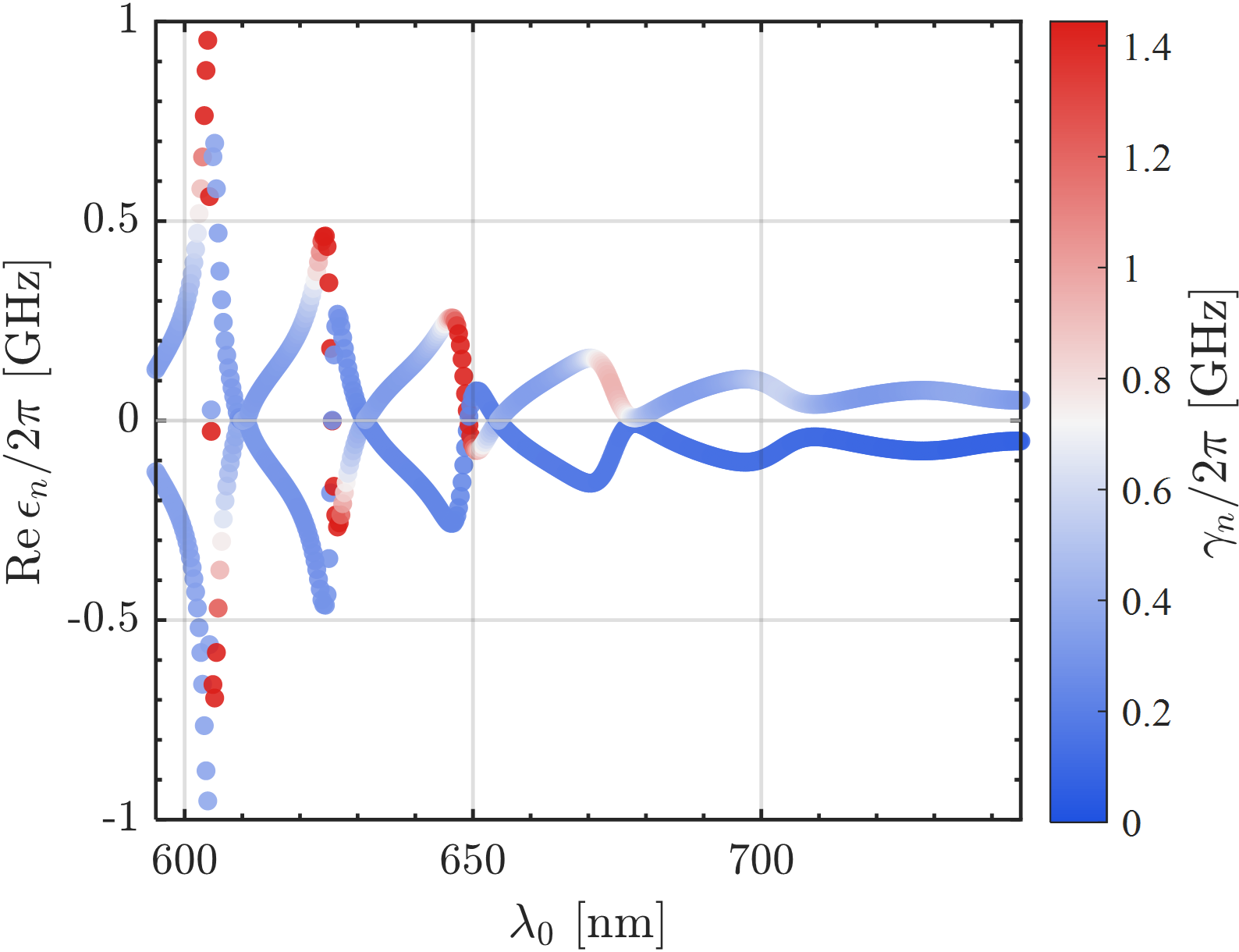}
    {\small (a)}
\end{minipage}

\begin{minipage}{0.85\linewidth}
    \centering
    \includegraphics[width=\linewidth]{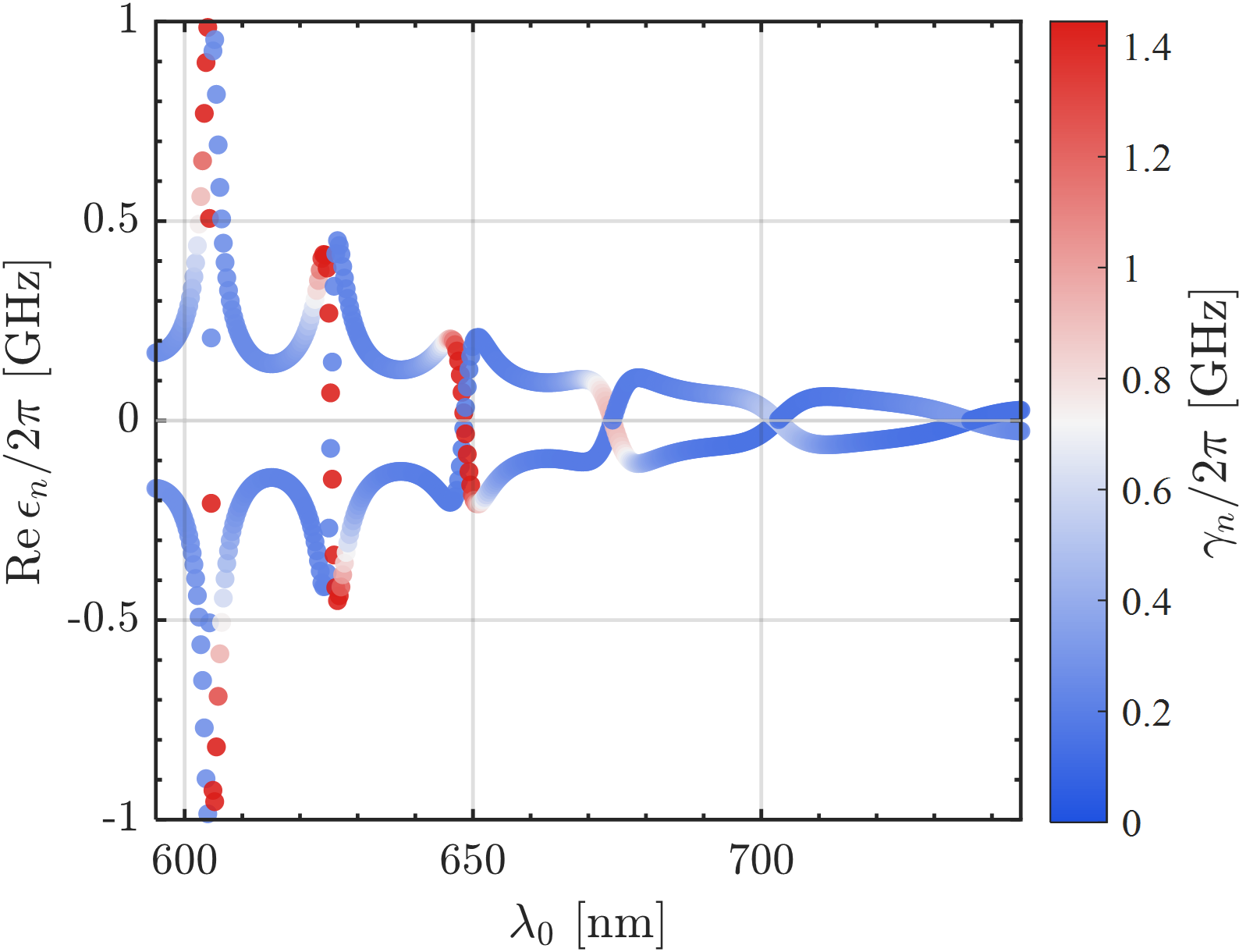}
    {\small (b)}
\end{minipage}

\caption{
Collective eigenmode spectra of the ring-resonator emitter configurations.
(a) Close two-emitter configuration.
(b) Distant two-emitter configuration.
The on-shell matrix
$\mathbf{H}_{\mathrm{eff}}=\mathbf{J}^{\mathrm{phys}}-i\pi\boldsymbol{\Gamma}$
is diagonalized for each $\lambda_0$.
The vertical coordinate gives $\mathrm{Re}\,\epsilon_n/2\pi$, while the color scale gives $\gamma_n/2\pi$.
}
\label{fig:dispersive_energy_splitting}
\end{figure}

Figure~\ref{fig:dispersive_energy_splitting} shows the collective eigenmode
spectra for the close [Fig.~\ref{fig:dispersive_energy_splitting}(a)] and
distant [Fig.~\ref{fig:dispersive_energy_splitting}(b)] two-emitter
configurations. In both cases the two branches split most strongly near the
ring resonances and approach the bare transition frequency away from them,
while the linewidth (color scale) rises in the same spectral regions where the
bright collective decay channel opens.
The close and distant configurations exhibit qualitatively similar spectral
structures, since in both cases the coherent exchange is mediated by the same
resonator modes: the Green's-function components share the ring's resonant
poles, while the emitter positions enter only through the modal residues. The
two maps therefore agree in their resonance positions but differ in the
magnitude of the splittings and linewidths, a distinction that is made
explicit in the population dynamics below.
The eigenmode map thus provides a frequency-domain interpretation of the
time-domain dynamics: large branch splittings mark enhanced coherent exchange,
while large-linewidth branches correspond to bright collective radiative
channels. This shows that the proposed full-wave Green's-function formulation
captures coherent energy shifts and collective radiative linewidths from a
single set of electromagnetic data.

\begin{figure}
\centering

% ---------- first row: Example 1 ----------
\begin{minipage}[t]{0.49\linewidth}
    \centering
    \includegraphics[width=\linewidth]{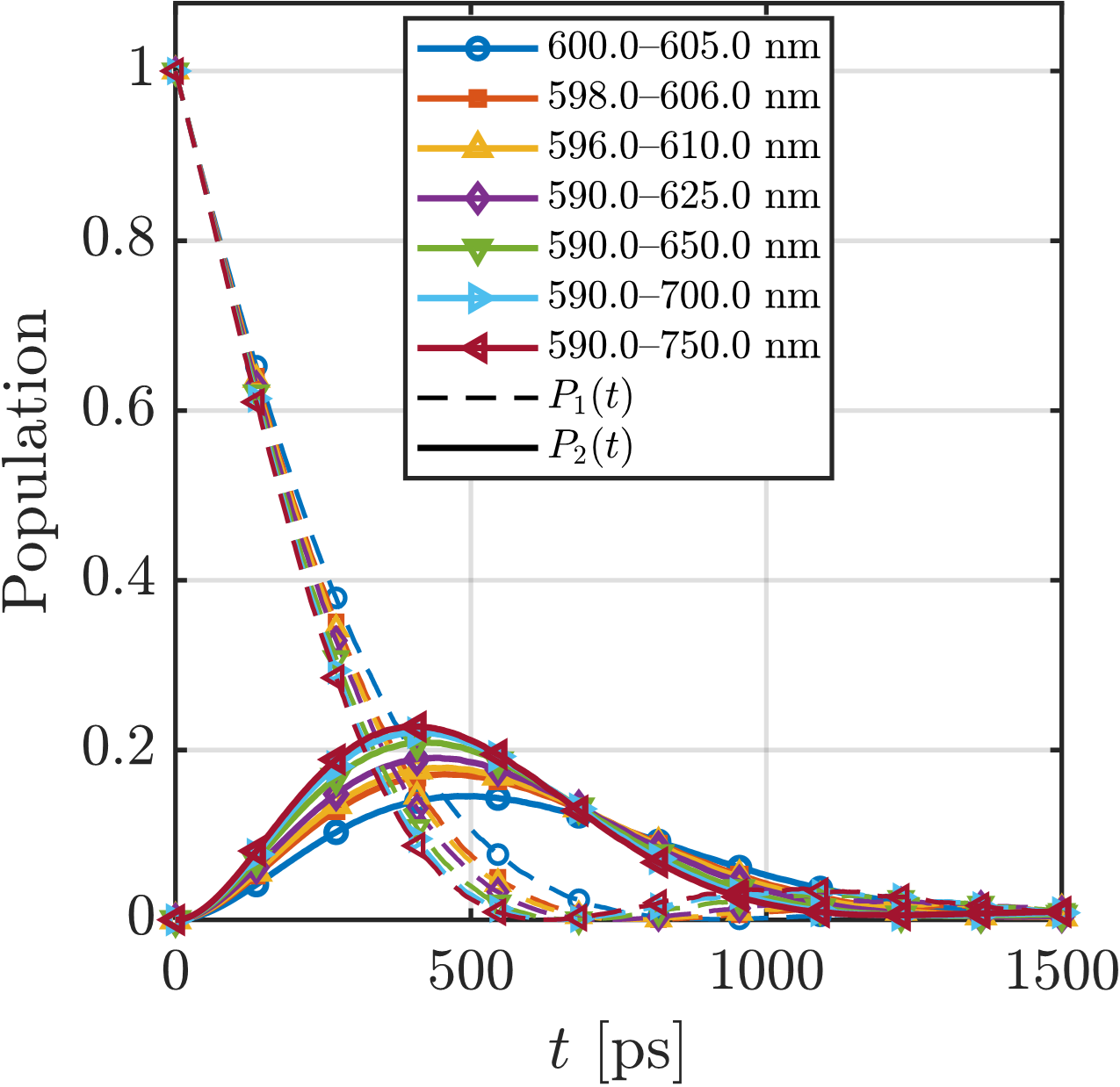}

    {\small (a)}
\end{minipage}
\begin{minipage}[t]{0.48\linewidth}
    \centering
    \includegraphics[width=\linewidth]{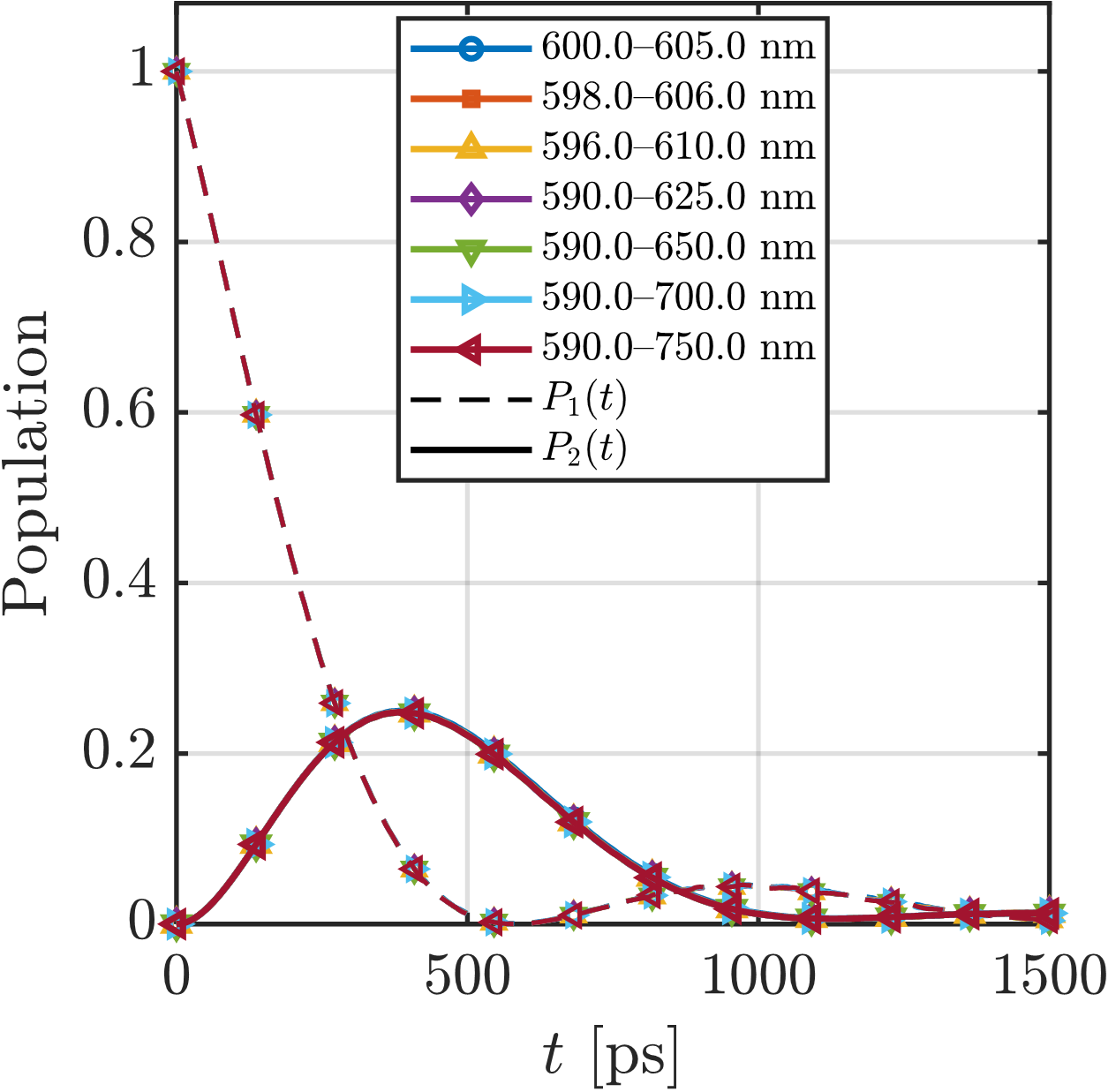}

    {\small (b)}
\end{minipage}

% ---------- second row: Example 2 ----------
\begin{minipage}[t]{0.49\linewidth}
    \centering
    \includegraphics[width=\linewidth]{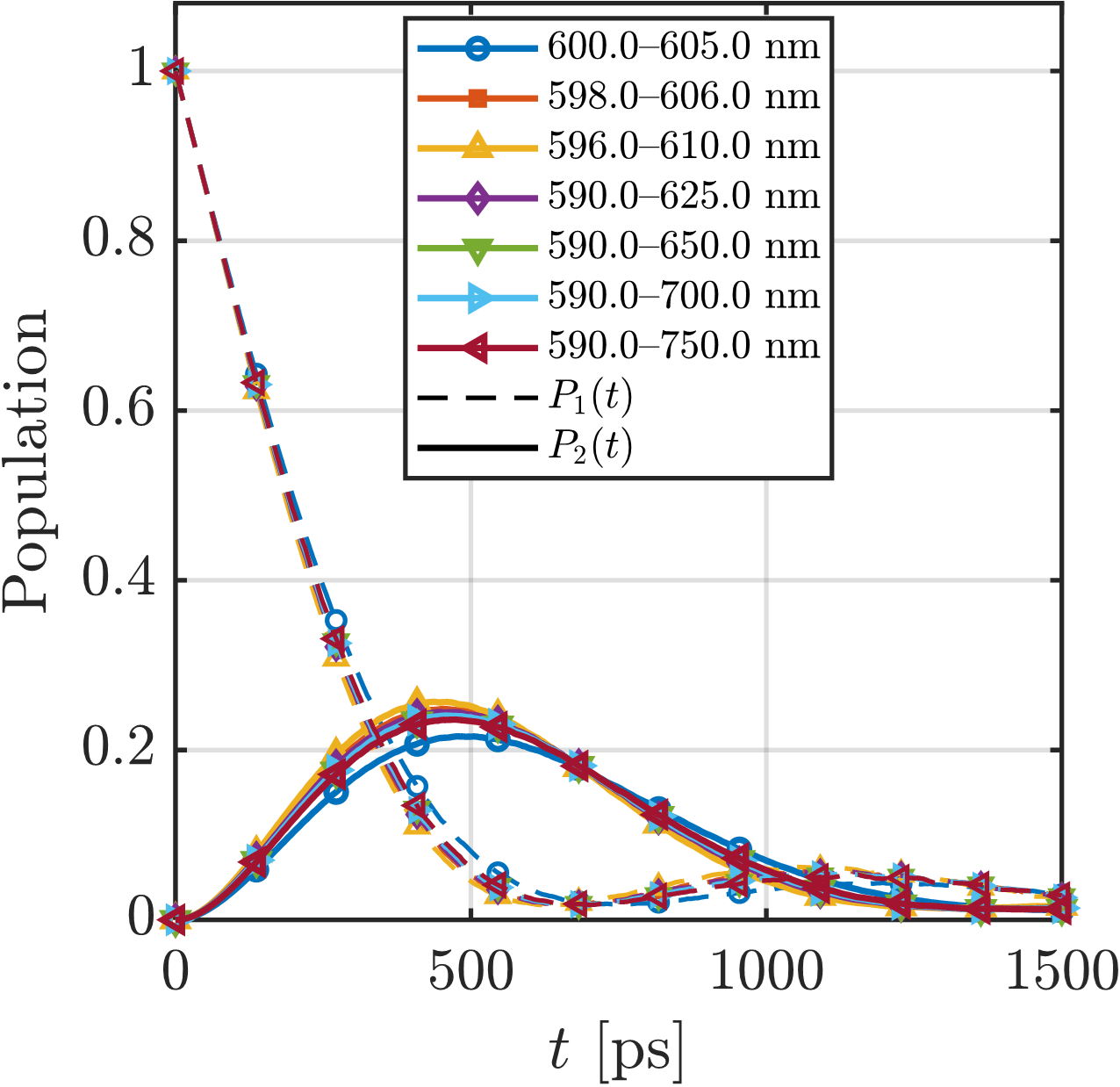}

    {\small (c)}
\end{minipage}
\hfill
\begin{minipage}[t]{0.48\linewidth}
    \centering
    \includegraphics[width=\linewidth]{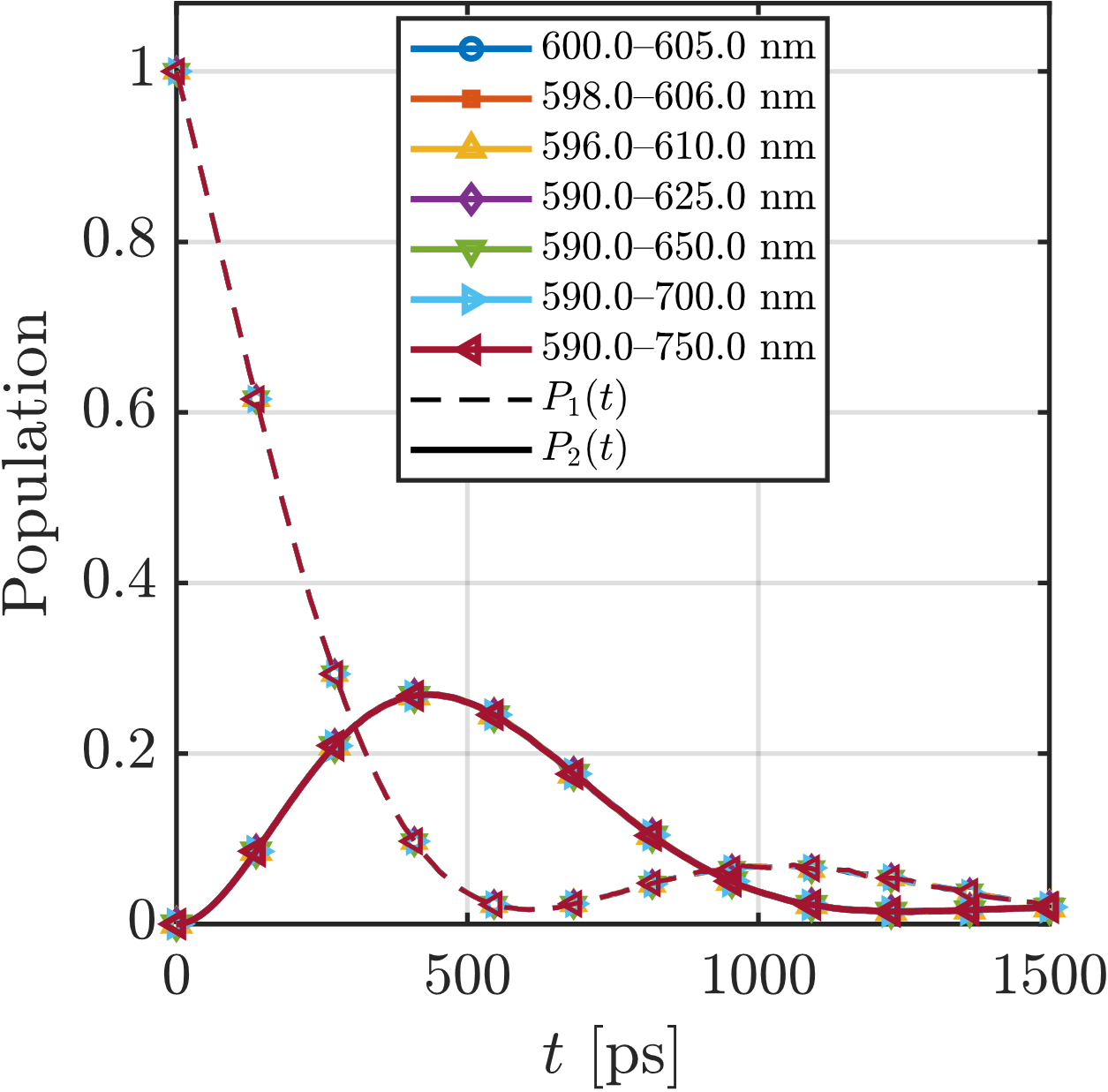}

    {\small (d)}
\end{minipage}

\caption{
Bandwidth-dependent population dynamics in the ring-resonator configurations.
(a) Uncorrected finite-band dynamics for the close two-emitter configuration.
(b) Corrected dynamics for the close two-emitter configuration after applying the counter-term $\delta J(B)=J^{\mathrm{phys}}-J^{\mathrm{trun}}(B)$.
(c) Uncorrected finite-band dynamics for the distant two-emitter configuration.
(d) Corrected dynamics for the distant two-emitter configuration after applying the same counter-term compensation.
Dashed and solid curves denote $P_1(t)$ and $P_2(t)$, respectively.
}
\label{fig:population_compensation}
\end{figure}

Figure~\ref{fig:population_compensation} demonstrates the role of the
compensation scheme in removing the artificial bandwidth dependence of the
coherent population dynamics. The emitter transition is tuned to
$\lambda_0=602~\mathrm{nm}$, away from the ring resonances, so that the finite-bandwidth bias is exposed most clearly. Both emitters carry a transition dipole moment $d_0=1.3\times10^{-28}~\mathrm{C\cdot m}$
oriented along the $z$-axis, with the system initialized as
$\mathbf C(0)=[1,0]^{T}$. Each panel superimposes the dynamics computed over
seven retained spectral windows, from $600$--$605~\mathrm{nm}$ to
$590$--$750~\mathrm{nm}$, for the close
[Figs.~\ref{fig:population_compensation}(a),(b)] and distant
[Figs.~\ref{fig:population_compensation}(c),(d)] two-emitter configurations.

In the uncorrected case [Figs.~\ref{fig:population_compensation}(a)
and~\ref{fig:population_compensation}(c)], the population-transfer dynamics
depends strongly on the retained bandwidth: as the window is widened, the
truncated principal-value exchange $J^{\mathrm{trun}}(B)$ drifts, and the
transfer curves spread into a visibly bandwidth-dependent family. This
sensitivity is more pronounced for the close pair
[Fig.~\ref{fig:population_compensation}(a)] than for the distant pair
[Fig.~\ref{fig:population_compensation}(c)], consistent with the stronger
near-field coupling of the closely spaced emitters. In either case, a bare
finite-band model does not provide a bandwidth-robust prediction of the
coherent dynamics, since the exchange interaction is contaminated by the
arbitrary spectral cutoff.
After the counter-term $\delta J(B)=J^{\mathrm{phys}}-J^{\mathrm{trun}}(B)$ is
applied [Figs.~\ref{fig:population_compensation}(b)
and~\ref{fig:population_compensation}(d)], the curves for all windows collapse
onto a nearly common trajectory. Because $J^{\mathrm{trun}}(B)+\delta J(B)$ is
locked to the on-shell physical exchange $J^{\mathrm{phys}}$ evaluated
independently from the real part of the dyadic Green's function, this trajectory
is bandwidth-robust while the explicitly propagated non-Markovian memory kernel
$\boldsymbol{\Gamma}(\omega)$ remains unchanged.

Having verified the compensation scheme in the off-resonant case, we now use
the compensated finite-band model with the full retained spectral window and
apply it at the ring resonance $\lambda_0=648.5~\mathrm{nm}$. The emitters keep the same dipole
moment $d_0=1.3\times10^{-28}~\mathrm{C\cdot m}$ along the $z$-axis, and we
consider the symmetric and antisymmetric initial states
\begin{equation}
\ket{\psi_{\pm}(0)}=\frac{1}{\sqrt{2}}\left(\ket{e_1,g_2}\pm\ket{g_1,e_2}\right),
\end{equation}
together with the localized state $[1,0]^{T}$.

\begin{figure}
\centering

\includegraphics[width=0.7\linewidth]{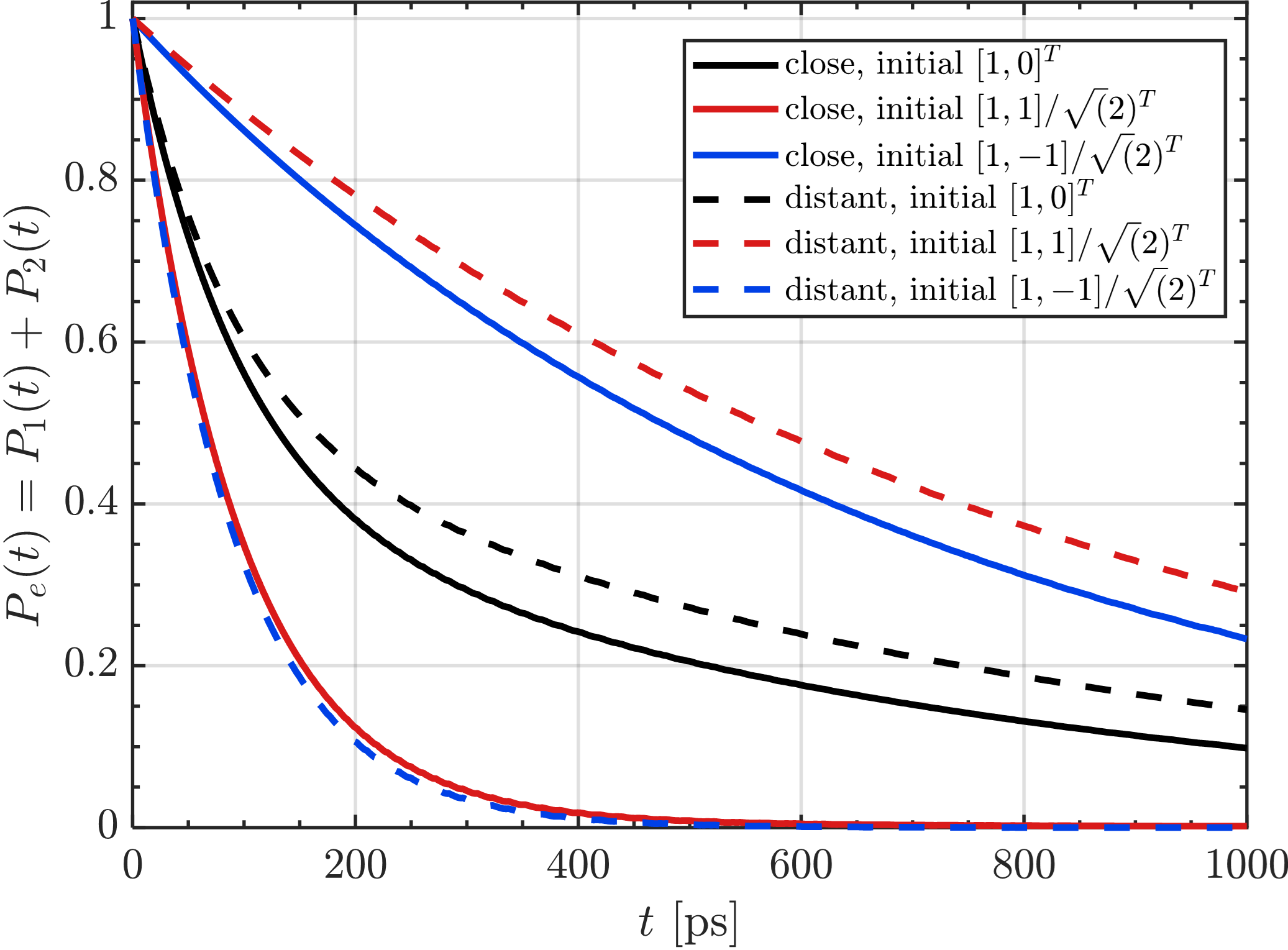}
\vspace{1mm}

{\small (a)}

\vspace{2mm}

\includegraphics[width=0.7\linewidth]{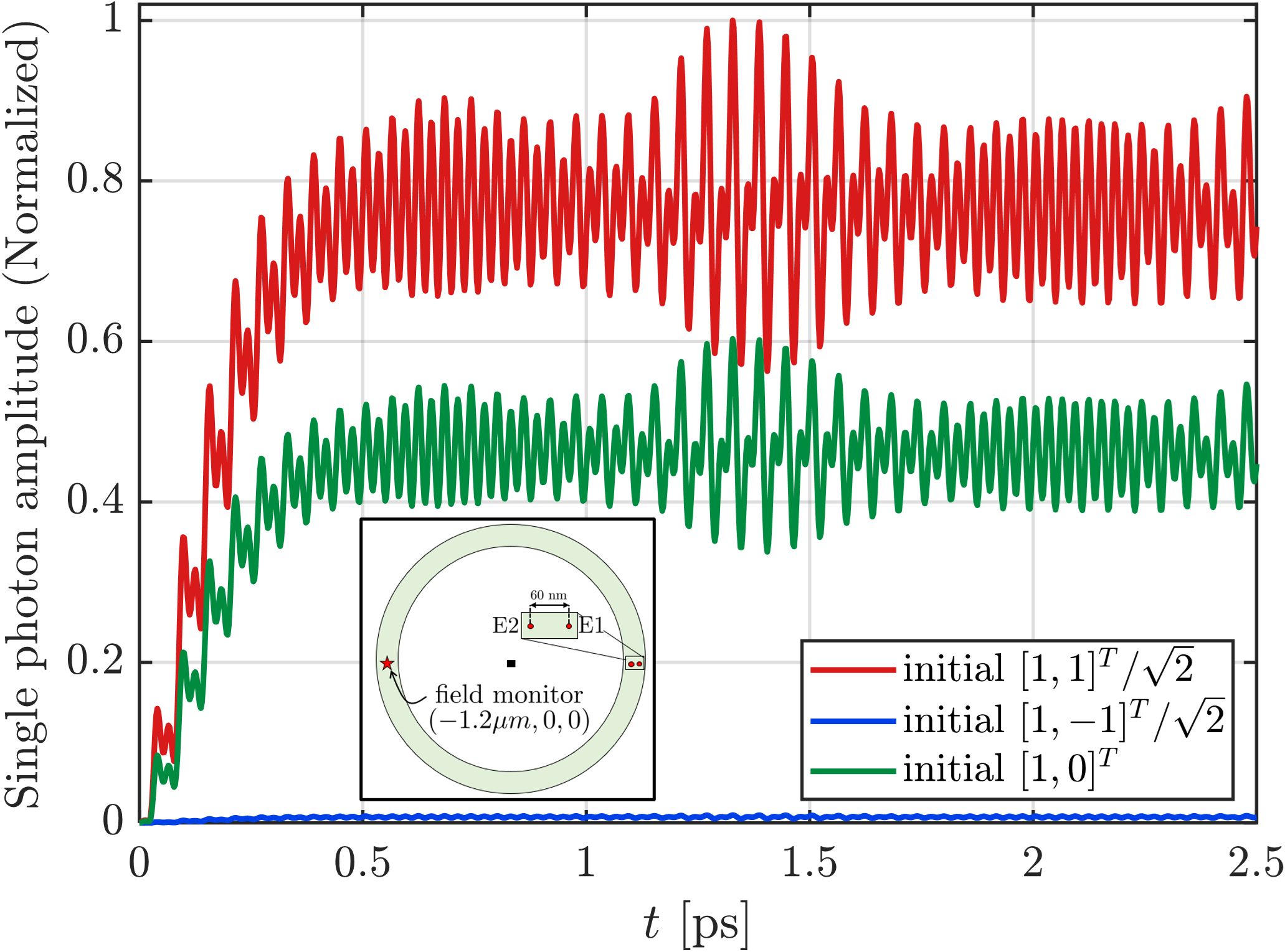}
\vspace{1mm}

{\small (b)}

\vspace{2mm}

\includegraphics[width=0.7\linewidth]{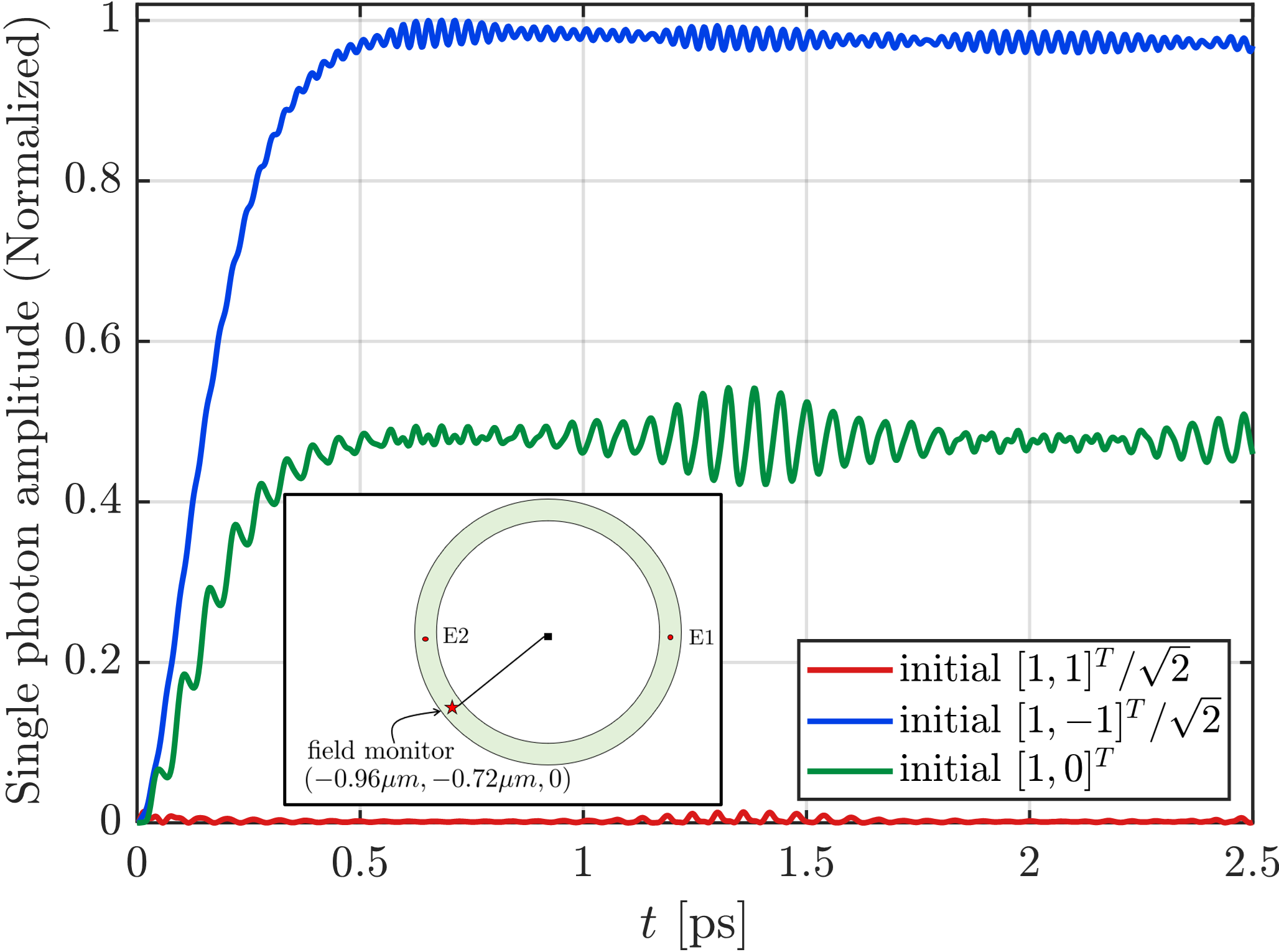}
\vspace{1mm}

{\small (c)}

\caption{
Full-spectrum dynamics at the ring resonance $\lambda_0=648.5~\mathrm{nm}$.
(a) Population dynamics for the close and distant two-emitter configurations.
(b),(c) Normalized single-photon amplitude for the close and distant
configurations, evaluated at off-emitter monitors
$\mathbf r_{\mathrm m}=(-1.2~\mu\mathrm{m},0,0)$ and
$(-0.96~\mu\mathrm{m},-0.72~\mu\mathrm{m},0)$, respectively.
}
\label{fig:population_spa_resonance}
\end{figure}
Figure~\ref{fig:population_spa_resonance}(a) shows the total excited-state
population $P_e(t)=P_1(t)+P_2(t)$ for the close (solid) and distant (dashed)
pairs. For the close pair the symmetric state is bright and decays rapidly while
the antisymmetric state is subradiant; for the distant pair this assignment is
reversed. The interchange follows from the sign of the off-diagonal collective
decay rate, set by $\mathrm{Im}\,G_{12}(\omega_0)$ for the close pair and
$\mathrm{Im}\,G_{13}(\omega_0)$ for the distant pair, so that the same resonator
supports opposite super- and subradiant superpositions depending on the emitter
separation. The localized state decays at an intermediate rate.
Figures~\ref{fig:population_spa_resonance}(b) and~(c) show the normalized
squared single-photon amplitude \( |E_{\mathrm{spa}}(\mathbf r_{\mathrm m},t)|^2 \)
at off-emitter monitors on the opposite side of the ring---$\mathbf r_{\mathrm m}=(-1.2~\mu\mathrm{m},0,0)$ for
the close pair and $(-0.96~\mu\mathrm{m},-0.72~\mu\mathrm{m},0)$ for the distant
pair. The emitted field mirrors the population dynamics: the bright
superposition---symmetric for the close pair, antisymmetric for the distant
pair---builds up a strong single-photon amplitude at the monitor, while the dark
one remains suppressed and the localized state gives an intermediate signal.
Since the framework retains the field amplitudes explicitly, these populations
and the emitted single-photon field follow from the same closed equations
without the quantum regression theorem, confirming that the proposed formulation
captures both collective decay and directional single-photon emission from a
single set of Green's-function data.

We now extend the analysis to a four-emitter array in Figure~\ref{fig:ring_geometries}(c). Four identical emitters with dipole moment
$d_0=1.3\times10^{-28}~\mathrm{C\cdot m}$ along the $z$-axis are placed at the
$C_4$-symmetric positions of Fig.~\ref{fig:ring_geometries}(c), at the ring
resonance $\lambda_0=648.5~\mathrm{nm}$.
Using $C_4$ symmetry, the dissipative kernel $\boldsymbol{\Gamma}(\omega_0)$ and the coherent
exchange $\mathbf{J}^{\mathrm{phys}}(\omega_0)$ take the circulant form
\begin{equation}
\mathbf{M}=
\begin{pmatrix}
M_{0} & M_{1} & M_{2} & M_{1}\\
M_{1} & M_{0} & M_{1} & M_{2}\\
M_{2} & M_{1} & M_{0} & M_{1}\\
M_{1} & M_{2} & M_{1} & M_{0}
\end{pmatrix},
\label{eq:circulant4}
\end{equation}
where $M_0$, $M_1$, and $M_2$ derive from $G_{11}$, $G_{12}$, and $G_{13}$,
respectively. Any circulant matrix is diagonalized by the discrete Fourier
basis, so the collective eigenstates of the on-shell Hamiltonian
$\mathbf{H}_{\mathrm{eff}}(\omega_0)=\mathbf{J}^{\mathrm{phys}}-i\pi\boldsymbol{\Gamma}$
are the angular-momentum-like superpositions
\begin{equation}
\ket{\psi_m}=\frac{1}{2}\sum_{j=1}^{4} e^{\,i\frac{\pi}{2} m (j-1)}\ket{e_j},
\qquad m=0,\pm1,2,
\label{eq:c4_eigenstates}
\end{equation}
independently of the numerical values of $M_0$, $M_1$, $M_2$. The corresponding
complex eigenvalues follow as $\epsilon_m=\Omega_m-i\gamma_m/2$, with
$\Omega_m$ the collective energy shift and $\gamma_m$ the collective decay rate;
these are obtained directly from the full-wave Green's-function data through
$\mathbf{J}^{\mathrm{phys}}$ and $\boldsymbol{\Gamma}$.

\begin{figure}
\centering

\begin{minipage}[t]{0.8\linewidth}
    \centering
    \includegraphics[width=\linewidth]{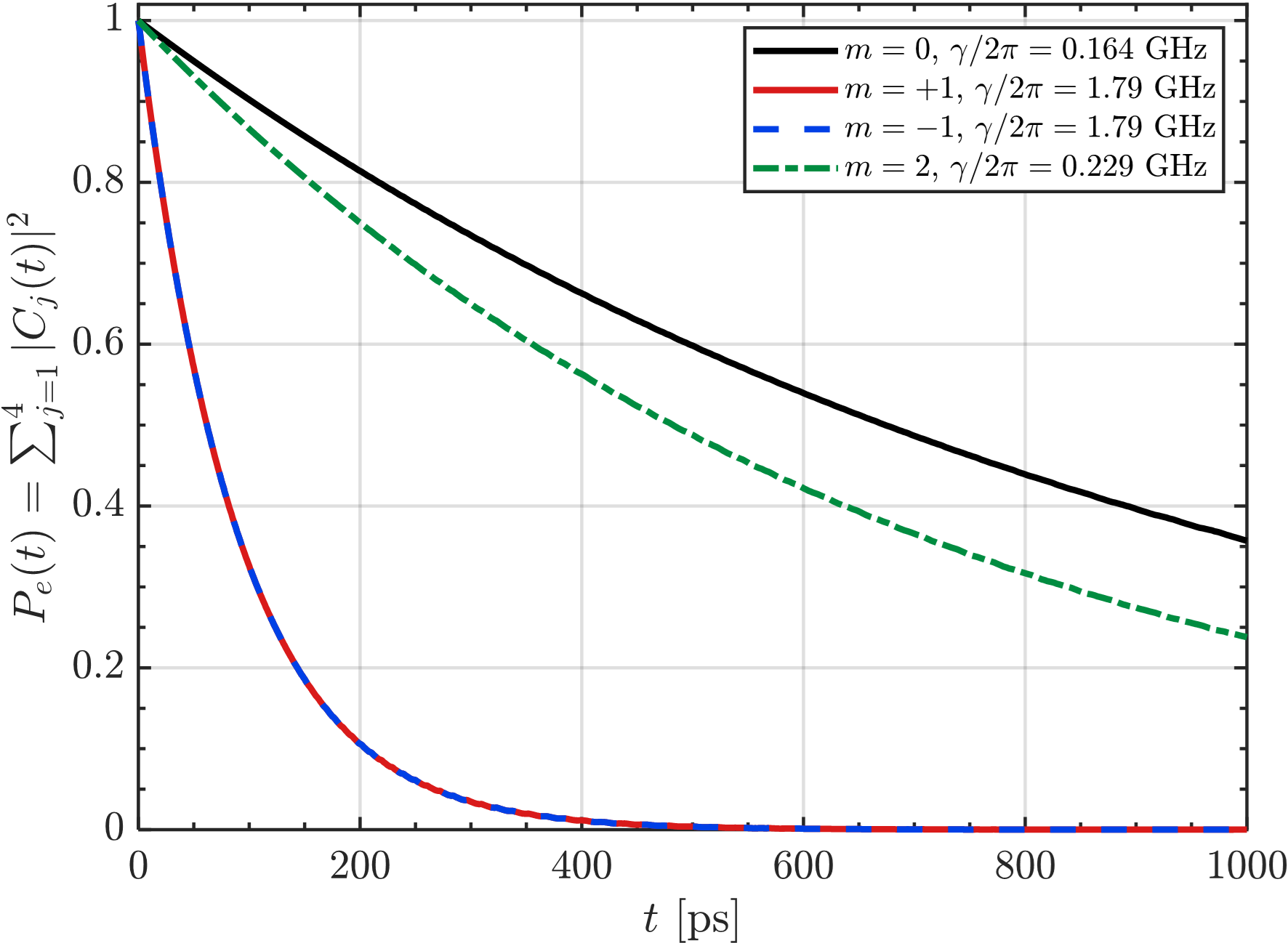}

    {\small (a)}
\end{minipage}

\begin{minipage}[t]{0.8\linewidth}
    \centering
    \includegraphics[width=\linewidth]{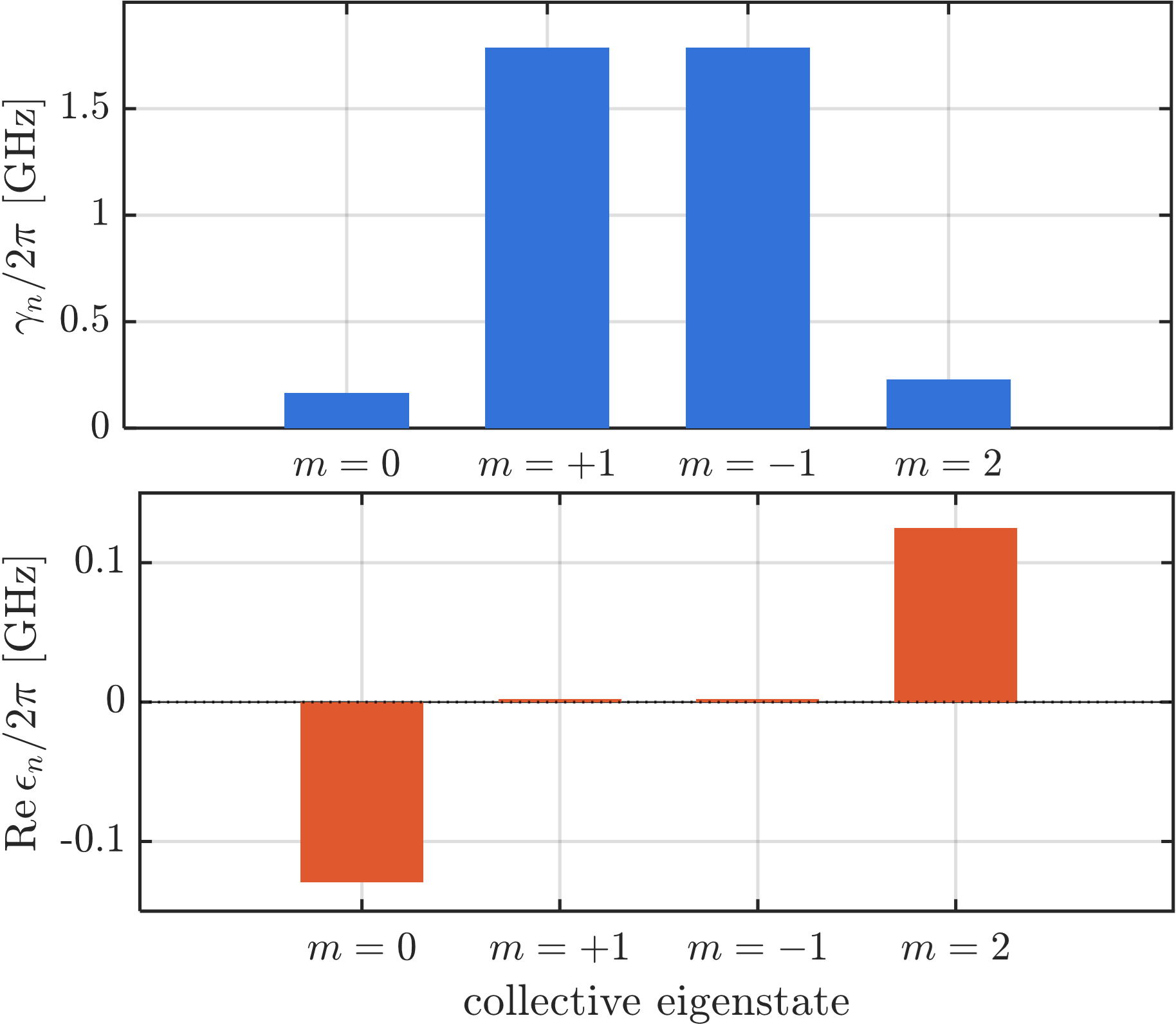}

    {\small (b)}
\end{minipage}

\caption{
Four-emitter collective dynamics at the ring resonance $\lambda_0=648.5~\mathrm{nm}$.
(a) Excited-state population decay for different angular-momentum-like collective eigenstates.
(b) Collective decay rates $\gamma_n/2\pi$ and coherent energy shifts
$\mathrm{Re}\,\epsilon_n/2\pi$ of the corresponding eigenstates.
}
\label{fig:four_emitter_resonant_dynamics}
\end{figure}

Figure~\ref{fig:four_emitter_resonant_dynamics}(a) shows the total excited-state
population $P_e(t)=\sum_{j=1}^{4}|C_j(t)|^2$ obtained by propagating each
collective eigenstate of Eq.~\eqref{eq:c4_eigenstates} using the compensated
finite-band model with the full retained spectral window. The degenerate $m=\pm1$ states decay most rapidly as bright superradiant
channels ($\gamma/2\pi\approx1.79~\mathrm{GHz}$), while $m=0$ and $m=2$ are
long-lived and subradiant ($\gamma/2\pi\approx0.16$ and $0.23~\mathrm{GHz}$,
respectively). The exact overlap of the $m=+1$ and $m=-1$ traces reflects the
degeneracy enforced by the $C_4$ symmetry: clockwise and counter-clockwise
circulating superpositions are mirror images and couple identically to the
rotationally symmetric resonator field.
Figure~\ref{fig:four_emitter_resonant_dynamics}(b) summarizes the collective
decay rates $\gamma_n/2\pi$ and coherent energy shifts
$\mathrm{Re}\,\epsilon_n/2\pi$ of the four eigenmodes. The bright $m=\pm1$ states
carry the largest decay rates but a vanishing collective shift, whereas the
subradiant $m=0$ and $m=2$ states acquire opposite-signed dispersive shifts with
much smaller decay rates. This complementary distribution of collective
linewidth and energy shift across the eigenmodes shows that the proposed
formulation resolves the full collective structure---both radiative and
dispersive---of a multi-emitter array directly from a single set of full-wave
Green's-function data.

\begin{figure}
\centering

\includegraphics[width=0.98\linewidth]{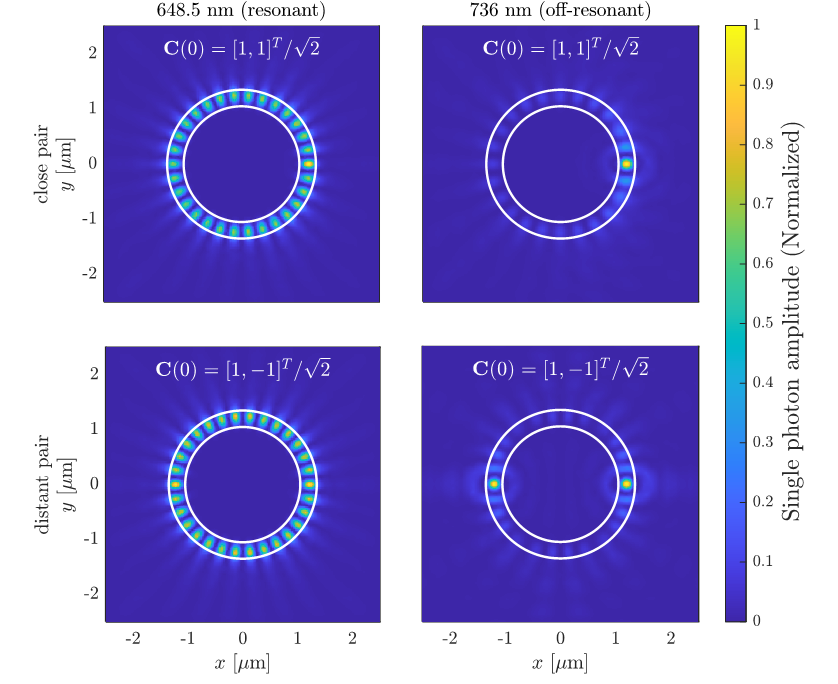}

\caption{
Two-emitter single-photon amplitude at $t=1~\mathrm{ps}$.
The normalized field distribution is compared for
$\lambda_0=648.5~\mathrm{nm}$ and $\lambda_0=736~\mathrm{nm}$.
The upper and lower rows correspond to the close and distant two-emitter configurations with
$\mathbf C(0)=[1,1]^T/\sqrt{2}$ and
$\mathbf C(0)=[1,-1]^T/\sqrt{2}$, respectively.
}
\label{fig:two_emitter_spa}
\end{figure}

Figure~\ref{fig:two_emitter_spa} shows the spatial distribution of the
single-photon amplitude at $t=1~\mathrm{ps}$ in the plane of the ring, for the
close (upper row) and distant (lower row) configurations, comparing the resonant
wavelength $\lambda_0=648.5~\mathrm{nm}$ with the off-resonant wavelength
$\lambda_0=736~\mathrm{nm}$. In each case the emitters are prepared in the bright
collective state of that geometry, and each panel is individually normalized to
its own maximum. We emphasize that the plotted quantity is the single-photon
amplitude defined in Eq.~\eqref{eq:E_spa}, not a classical field intensity.
It is the projection of the emitted one-photon state onto the field vacuum and
therefore represents the genuinely quantum-mechanical amplitude at position $\mathbf r$, reconstructed directly from the
frequency-resolved amplitudes $E_\omega^{(p)}(t)$ through
Eq.~\eqref{eq:spa_multi_final} without invoking the quantum regression theorem.

At the resonance, the single-photon amplitude organizes into a clear
resonator-mode pattern delocalized around the entire ring waveguide, showing that
the bright collective state emits its photon into the resonator-supported mode.
At the off-resonant wavelength, by contrast, the amplitude remains localized near
the emitters and does not develop into a ring-spanning mode pattern. This spatial
contrast complements the time-domain dynamics of
Fig.~\ref{fig:population_spa_resonance}: the same bright collective channel that
decays rapidly in the population dynamics is the one whose single-photon amplitude
is coherently distributed over the resonator mode. That the full spatiotemporal
structure of the emitted single-photon field follows from the same
Green's-function data as the atomic populations underscores the central capability
of the proposed framework---the simultaneous, first-principles modeling of emitter
dynamics and structure-resolved single-photon emission in an open, dissipative
electromagnetic environment.

\begin{figure*}
\centering

\includegraphics[width=0.9\textwidth]{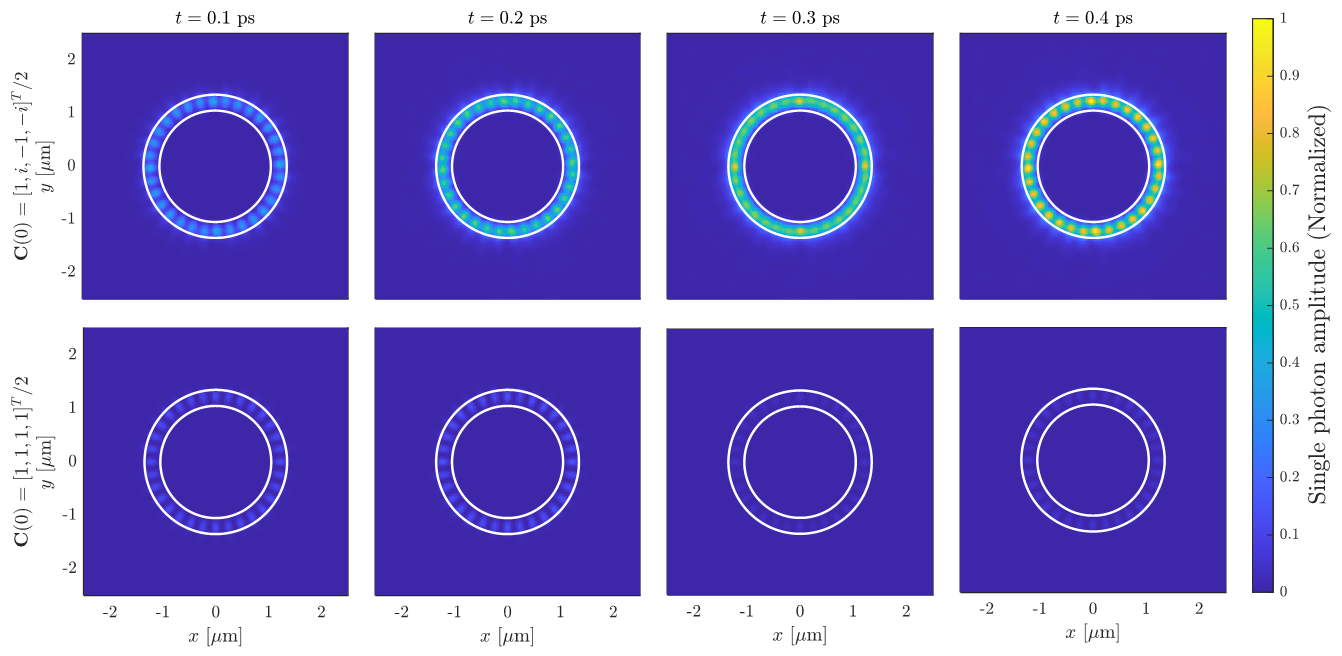}

\caption{
Four-emitter single-photon amplitude at the ring resonance
$\lambda_0=648.5~\mathrm{nm}$.
The normalized spatial distribution of the single-photon amplitude is shown at
$t=0.1$, $0.2$, $0.3$, and $0.4~\mathrm{ps}$.
The upper row corresponds to the initial collective state
$\mathbf C(0)=[1,i,-1,-i]^T/2$,
whereas the lower row corresponds to
$\mathbf C(0)=[1,1,1,1]^T/2$.
The white curves indicate the inner and outer boundaries of the
$\mathrm{Si}_3\mathrm{N}_4$ ring resonator.
}
\label{fig:four_emitter_spa}
\end{figure*}

Figure~\ref{fig:four_emitter_spa} shows the time evolution of the four-emitter
single-photon amplitude at the ring resonance $\lambda_0=648.5~\mathrm{nm}$,
displayed as spatial snapshots of $|E_{\mathrm{spa},z}|^2$ at $t=0.1$, $0.2$,
$0.3$, and $0.4~\mathrm{ps}$. The upper row is initialized in the bright
collective eigenstate $\mathbf{C}(0)=[1,i,-1,-i]^{T}/2$ ($m=+1$), and the lower
row in the subradiant eigenstate $\mathbf{C}(0)=[1,1,1,1]^{T}/2$ ($m=0$),
following the mode assignment of Fig.~\ref{fig:four_emitter_resonant_dynamics}.
As in the two-emitter case, the plotted quantity is the quantum single-photon
amplitude of Eq.~\eqref{eq:E_spa}, reconstructed from the same full-wave
Green's-function data.
For the bright $m=+1$ state, the single-photon amplitude progressively builds up
into a ring resonator wave pattern that wraps around the entire waveguide,
visualizing in real time how the superradiant collective state emits its photon
into the delocalized resonator mode. For the subradiant $m=0$ state, the
amplitude remains weak and confined near the emitters throughout the same time
window, consistent with its strongly suppressed collective decay rate. This
time-resolved spatial map thus directly connects the collective decay rates of
Fig.~\ref{fig:four_emitter_resonant_dynamics} to the emitted single-photon field,
confirming that the proposed framework resolves the full spatiotemporal structure
of multi-emitter collective emission in a realistic three-dimensional resonator
from a single set of calculated Green's-function data.

\section{Conclusions}

This work establishes a Green's-function-based route for treating collective
single-photon emission in open, dispersive, and lossy electromagnetic
environments without reducing the structure to a small set of phenomenological
cavity modes. By combining the modified Langevin noise formalism with a
frequency-resolved single-excitation description, the electromagnetic
environment is represented directly through the full-wave dyadic Green's
function. The resulting equations retain the non-Markovian reservoir amplitudes
explicitly, so that the emitter populations and the emitted single-photon field
are obtained within the same dynamical calculation.

A key outcome of this formulation is the identification of a practical
consistency issue that arises when numerically evaluated Green's functions are
used over a finite spectral window. The dissipative memory kernel may appear
well converged within the retained bandwidth, while the corresponding coherent
exchange interactions remain biased because their principal-value
reconstruction depends on spectral components outside the simulated window.
This separation between dissipative convergence and dispersive convergence is
especially relevant in structured photonic environments, where near-field
couplings and resonant modal tails can strongly affect the real part of the
Green's function. The counter-term compensation scheme introduced here resolves
this issue by restoring the physical on-shell dispersive interaction from the
real part of the Green's function, without altering the explicitly propagated
finite-band memory kernel.

The numerical examples illustrate how this correction changes the predictive
content of full-wave multi-emitter simulations. In free space, the compensation
recovers the analytic coherent exchange interaction. In the plasmonic
nanosphere example, it separates the convergence of the on-shell dissipative
response from the residual dispersive bias. In the dielectric ring resonator,
the same framework captures bandwidth-robust population transfer, collective
bright and subradiant channels, and the spatial build-up of the emitted
single-photon amplitude in realistic three-dimensional geometries. These
results show that full-wave electromagnetic data can be used not only to compute
decay rates or Purcell factors, but also to propagate collective quantum
dynamics and structure-resolved single-photon emission in a unified way.

The present formulation is restricted to the single-excitation manifold, but it
provides a basis for several extensions. Future work may incorporate
multi-excitation dynamics, larger and disordered emitter arrays, time-dependent
or nonlinear material responses, and inverse-designed photonic structures.
Because the method connects electromagnetic design parameters directly to
non-Markovian quantum dynamics and emitted single-photon waveforms, it may be
useful for the optimization of deterministic single-photon sources, collective
emitter arrays, and coherent quantum interfaces in realistic nanophotonic
platforms.

\bibliography{sorsamp}

\newpage
 
\end{document}